# The quest to quantify selective and synergistic effects of plasma for cancer treatment: Insights from mathematical modeling

**Charlotta Bengtson * and Annemie Bogaerts**

Research Group PLASMANT, Department of Chemistry, University of Antwerp, Universiteitsplein 1, B-2610 Wilrijk-Antwerp, Belgium
***** Correspondence: charlotta.bengtson@uantwerpen.be

**Abstract:** Cold atmospheric plasma (CAP) and plasma-treated liquids (PTLs) have recently become a promising option for cancer treatment, but the underlying mechanisms of the anti-cancer effect are still to a large extent unknown. Although hydrogen peroxide ($H_2O_2$) has been recognized as the major anti-cancer agent of PTL and may enable selectivity in a certain concentration regime, the co-existence of nitrite can create a synergistic effect. We develop a mathematical model to describe the key species and features of the cellular response towards PTL. From the numerical solutions, we define a number of dependent variables, which represent feasible measures to quantify cell susceptibility in terms of the $H_2O_2$ membrane diffusion rate constant and the intracellular catalase concentration. For each of these dependent variables, we investigate the regimes of selective versus non-selective, and of synergistic versus non-synergistic effect to evaluate their potential role as a measure of cell susceptibility. Our results suggest that the maximal intracellular $H_2O_2$ concentration, which in the selective regime is almost four times greater for the most susceptible cells compared to the most resistant cells, could be used to quantify the cell susceptibility towards exogenous $H_2O_2$. We believe our theoretical approach brings novelty to the field of plasma oncology, and more broadly, to the field of redox biology, by proposing new ways to quantify the selective and synergistic anti-cancer effect of PTL in terms of inherent cell features.

## 1. Introduction

In the last decade, the use of cold atmospheric plasma (CAP) - which is an ionized gas near room temperature - has become a novel method to treat cancer. Both direct application of CAP, e.g. by the clinically approved kINPenMED® plasma jet, and indirect treatment by application of plasma-treated liquids (PTLs), have been shown to provide a significant anti-cancer effect [1]. Van Boxem et al. [2] showed that PTLs have anti-cancer effect for a number of different CAP and liquid conditions, and Lin et al. [3] found that CAP can induce immunogenic cancer cell death. This mode of cell death induced by CAP was later attributed to the CAP generated short-lived reactive species [4]. Moreover, CAP and PTLs have been reported to cause a selective anti-cancer effect [5], although selectivity depends on the cell type, the type of cancer and the culturing medium [6]. Bekeschus et al. [7], demonstrated, using an *in ovo* model, that CAP is a safe cancer treatment modality with respect to possible metastasis formation. A number number of promising results of clinical application of CAP for cancer treatment have also been published, see e.g. refs. [8, 9].

It is widely believed that the processes leading to cancer cell death are initiated by reactive oxygen and/or nitrogen species (RONS), in particular hydrogen peroxide ($H_2O_2$), but the knowledge about the specific mechanisms underlying cell death induced by CAP and PTL is still very limited. The lack of understanding of the combined effect of RONS contained in CAP and PTL in terms of the cellular response to exposure, is problematic in the development of CAP/PTL treatment as a standardized cancer therapy for clinical use. Ultimately, it should be possible to predict and quantify the susceptibility to CAP/PTL of a particular cell line in terms of features specifically associated with those cells.

So far, the vast majority of the literature in plasma oncology are experimental studies. As a complement, another approach to increase the understanding of complex biological systems, such as the interaction between cells and PTL, is to develop a mathematical model that includes all the known information (of major importance in the given context) about the system and use it to investigate the system's response to various conditions. Especially, the system's response to a perturbation of the "normal" conditions can be analyzed. Furthermore, the development of the mathematical model itself can be seen as a way to summarize the current state of knowledge on the matter in a compact manner; it can be seen as the current "working hypothesis" of the mechanisms and processes governing the system dynamics.

Mathematical modeling has indeed proven to be a useful approach to increase our knowledge about the mechanisms of the cell's antioxidant defense and redox signaling. Some examples are the range of diffusion of $H_2O_2$ in the cytosol [10, 11], and the cellular decomposition of exogenous $H_2O_2$ [12-14]. In the context of plasma oncology, two catalase-dependent apoptotic pathways associated with cancer cells, which possibly could be reactivated by CAP and thus explain anti-cancer effect of CAP, have been investigated by mathematical modeling [15]. It was found that these pathways are unlikely to account for the anti-cancer effect of CAP and thus the underlying cause has to be studied further.

In the present study, we develop a mathematical model that includes the species and mechanisms of major importance in the context of a cell system exposed to PTL. The ultimate aim is to find a measure in terms of key features and characteristics of cells, which is able to quantify a particular cell system's *susceptibility* towards PTL and thus explain differences in response between normal cells and cancer cells. To the best of our knowledge, this is a completely novel approach in the field of plasma oncology, and we believe that our study will provide a new perspective, and new insights, as a complement to experimental studies. An extensive summary of the background, leading to the more detailed research question, is provided in section 2.

## 2. Experimentally observed cytotoxic effects of CAP and PTL and possible features determining cancer cell susceptibility

An immediate effect of CAP treatment of cancer cells is an increase of intracellular RONS [16-20]. The significance of this RONS accumulation has been verified by the observation that the treatment does not succeed if the cancer cells have been pre-treated with intracellular RONS scavengers [17, 21, 22]. The origin of the increase in intracellular RONS after CAP treatment is still under investigation, but a hypothesis consistent with experimental observations is that it is caused by a diffusion of extracellular CAP-originated RONS across the cell membrane [17, 18, 23, 24].

It has been demonstrated that the anti-cancer effect of CAP can also be induced by the species in PTL. In PTL - which is mainly consisting of $H_2O_2$, $NO_2^-$ and $NO_3^-$ [25-27] - $H_2O_2$ has been shown to be of major importance [2, 25-31]. It has e.g. been demonstrated that the $H_2O_2$ consumption rate - which is cell specific - of cancer cells after PTL treatment, is a key factor determining the specific susceptibility of cancer cell lines to PTL. More explicitly, it has been reported that the higher $H_2O_2$ consumption rate of cancer cells, the lower is the susceptibility towards CAP/PTL [32]. The susceptibility of cancer cells towards exogenous $H_2O_2$ has also been shown in e.g. refs. [33-36]. However, it has been found that $H_2O_2$ alone cannot account for the total anti-cancer effect observed for PTL [27]. In this context, there are some reports of a synergistic effect of $H_2O_2$ and $NO_2^-$ in PTL [25, 26]. Thus, the cytotoxic effect of $H_2O_2$ seems to be enhanced in the presence of $NO_2^-$. Ref. [26] found a *selective, synergistic* anti-cancer effect for $H_2O_2$ in the $\mu M$-range and $NO_2^-$ in the $mM$-range, whereas in ref. [25] a *non-selective, synergistic* anti-cancer effect was reported when $H_2O_2$ and $NO_2^-$ were both in the $mM$-range. Since $H_2O_2$ and $NO_2^-$ in PTL may react to form $ONOO^-$ [37], which is known to be highly toxic to cells, it has been speculated whether $ONOO^-$ is the species causing the synergistic effect of $NO_2^-$ and $H_2O_2$. Its formation could thus potentially increase the cytotoxicity of PTL compared to an equal concentration of $H_2O_2$ only.

To summarize, some key points of the observed cytotoxic effects of CAP or PTL, are:

- An intracellular increase of RONS, which is likely to be caused by diffusion of CAP-originated constituents through the cell membrane, is crucial for cell cytotoxicity.

- The key species in the anti-cancer effect of PTL is $H_2O_2$ (note that this may be different for direct CAP treatment, where short-lived RONS also play a crucial role [4, 38]), and the corresponding cytotoxicity is inversely proportional to the extracellular consumption rate of $H_2O_2$.
- The effect of extracellular $H_2O_2$ is enhanced in the presence of $NO_2^-$, which can be a clue to understand why PTL enables a more efficient treatment than a mock solution of $H_2O_2$ only.

Thus, from the information presented in literature, we can conclude that the cellular response to an addition of extracellular $H_2O_2$ - with and without a simultaneous addition of $NO_2^-$ - is crucial to understand the anti-cancer effect of PTL. In section 2.1, we introduce the key parameters to predict the response towards extracellular $H_2O_2$ of cells. We also relate this to general differences between normal cells and cancer cells. Based on this information and knowledge, we introduce our approach, and formulate our research question and aim in detail, in section 2.2.

### 2.1. Differences in cellular response to exogenous hydrogen peroxide

Especially two factors determine whether a particular cell line is susceptible to exposure of exogenous $H_2O_2$:

- The plasma-membrane $H_2O_2$ diffusion rate constant
- The intracellular expression of catalase

Several cancer cell lines have shown a common phenotype of *decreased* catalase expression and *increased* aquaporin expression (that facilitates the transport of $H_2O_2$ through the cell membrane [39-42] and thus determines the $H_2O_2$ membrane diffusion rate) as compared to normal cells. Hence, cancer cells in general can be assumed to be more susceptible to exogenous $H_2O_2$.

#### *2.1.1. Membrane diffusion rate of hydrogen peroxide in normal versus cancer cells*

Aquaporins are proteins that form pores in the cell membrane. Primarily, they facilitate the transport of water between cells, but they also enable the trans-membrane diffusion of $H_2O_2$ (due to the chemical similarities between both molecules). Thus, the aquaporin expression in the cell membrane relates to the membrane diffusion rate of $H_2O_2$. Many aquaporins have been found to be over-expressed in tumors of different origins, especially in aggressive tumors [43]. Since different cancer cell lines express aquaporins to various extent [43, 44], the different responses of $H_2O_2$-exposure by different cancer cell lines can at least partly be explained by the non-identical levels of aquaporin expression. In e.g. ref. [45] it was found that aquaporin 3 accounts for nearly 80% of the membrane diffusion of $H_2O_2$ in a human pancreatic cancer cell line. For cells with a decreased aquaporin 3 expression, the rate of $H_2O_2$-uptake from the extracellular compartment was significantly decreased. It has furthermore been shown that for glioblastoma tumor cells, the anti-cancer effect of PTL - as well as the increase of the concentration of intracellular RONS - was significantly inhibited when aquaporin 8 was inhibited [46].

#### *2.1.2. Catalase activity in normal versus cancer cells*

Catalase is one of the main enzymes of the antioxidant defense system of cells of almost all aerobic organisms. The biological role of catalase is to regulate intracellular steady-state concentrations of $H_2O_2$, and experimental investigations and kinetic models using *in vitro* data have demonstrated that catalase is the major enzyme involved in the antioxidant defense against high concentrations of $H_2O_2$ [12, 47-49]. In particular, catalase has been shown to be responsible for the clearance of *exogenous* $H_2O_2$ *in vitro* and *in vivo* [12, 50-52].

Although catalase levels vary widely across cell lines, the total concentration of catalase (extracellular and intracellular) is frequently reported to be lower in cancer cells than in normal cells [36, 53-60]. In ref. [61] it was found that the catalase activity in various cancer cells is up to an order of magnitude lower compared to normal cells, and in ref. [62] it was shown that normal cells have a better capacity to remove extracellular $H_2O_2$ than cancer cells; the rate constants for removal of extracellular $H_2O_2$ were on average two times higher in normal cells than in cancer cells. Furthermore, it was reported in ref. [62] that the rate constants for $H_2O_2$ removal by different cell lines correlated with the number of active catalase monomers per cell.

However, while in general, the levels of catalase are low in cancer cells, catalase activity appears to vary greatly across different cancer cell lines [63]. In ref. [34], it was found that three cancer cell lines (glioblastoma) that were extremely susceptible to $H_2O_2$ (generated by ascorbic acid) had reduced activity of intracellular catalase. Ascorbic acid-resistant cancer cell lines, on the other hand, exhibited significantly higher levels of catalase, but catalase knockdown sensitized these cell lines to extracellular $H_2O_2$.

An additional aspect of catalase that may be of interest in the context of cytotoxicity of CAP and PTL, is that it has been shown to decompose $ONOO^-$ [64]. Thus, if the synergistic effect of $H_2O_2$ and $NO_2^-$ is to be found in the formation of $ONOO^-$, catalase might have a double function, i.e., as a protective factor towards exogenous exposure of both $H_2O_2$ and $NO_2^-$.

### 2.2. Approach and research question

In this study, we develop a mathematical model of the kinetics of the key species of PTL, i.e., $H_2O_2$ and $NO_2^-$, as well as of the processes governing the interaction with a cell system, which are given in terms of the $H_2O_2$ membrane diffusion rate constant and the intracellular catalase concentration. The system modeled is illustrated in Figure 1.

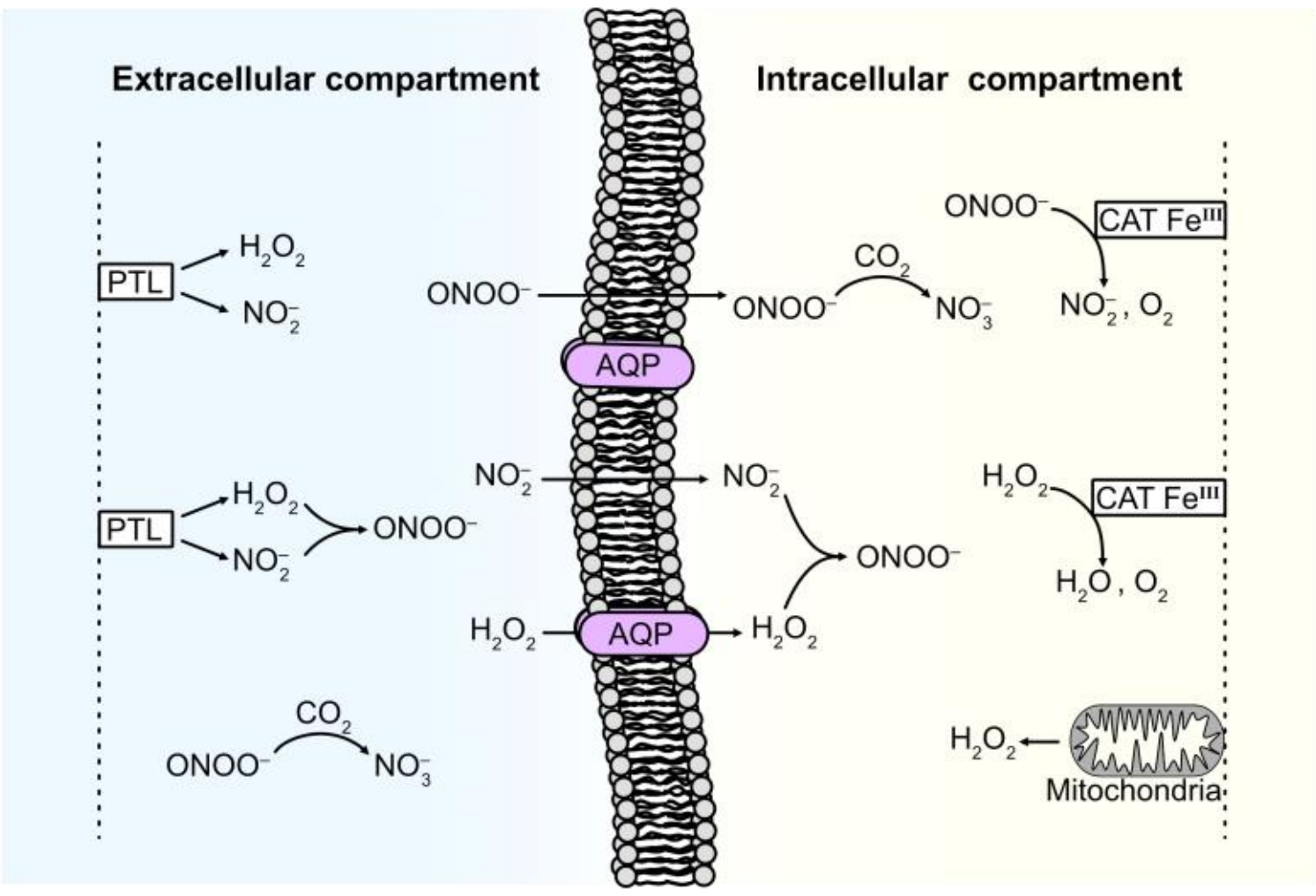

**Figure 1.** Illustration of the system representing a cell exposed to PTL.

As can be seen, the system consists of two compartments: the extracellular compartment (EC), and the intracellular compartment (IC). The two compartments are separated by the cell membrane, which some species in the system can diffuse through. Our mathematical model is explained in detail in section 5, with references to all input data and assumptions made. Briefly, it takes into account (i) the diffusion of $H_2O_2$ and $NO_2^-$ from the EC (where these species are supplied by PTL) to the IC, (ii) the formation of $ONOO^-$ from $H_2O_2$ and $NO_2^-$ (in both the IC and EC), (iii) the mitochondrial production of $H_2O_2$ (in the IC), and (iv) the decomposition of $H_2O_2$ and $ONOO^-$ (in the IC). Furthermore, (v) since the $CO_2$-catalyzed consumption is considered to be the main route for $ONOO^-$-decay in biological systems (due to a high $CO_2$-concentration) [65-69], this reaction is also included.

There have already been attempts to capture the susceptibility towards exogenous $H_2O_2$ of different cell lines in terms of their $H_2O_2$ membrane diffusion rate constant and intracellular catalase concentration [13, 14]. Two dependent variables that have been investigated recently are the intracellular steady-state concentration of $H_2O_2$ and the so-called latency (which describes the reduced average reaction rates for the observed decomposition of $H_2O_2$ due to the localization of encapsulated catalase in the peroxisomes). In ref. [13] a lumped-parameter mathematical model, assuming that catalase is the major $H_2O_2$-removal enzyme, was developed and used to calculate the intracellular steady-state $H_2O_2$ concentration for several cell lines. The model was calibrated to the experimental values of measured critical parameters, and the resulting intracellular steady-state $H_2O_2$ concentration was related to observed cell specific susceptibility to extracellular exposure of $H_2O_2$. The results showed that despite the fact that the experimental parameters, including catalase concentration and $H_2O_2$ membrane diffusion rate constant in particular, varied significantly across cell lines, the calculated steady-state intracellular-to-extracellular $[H_2O_2]$ ratio did not vary significantly across cell lines. In ref. [14], it was investigated whether variations in the latency of peroxisomal catalase across cancer cell lines correlates with observed *in vitro* susceptibility to ascorbate at equivalent dosing of extracellular $H_2O_2$. The so-called effectiveness factor - which takes both the membrane diffusion rate and the overall reduced activity for encapsulated catalase into account - was used to quantify the effect of latency. The results suggest that latency alone is not a reliable parameter for predicting cell susceptibility to ascorbate (and hence, $H_2O_2$).

In this study, we explore new dependent variables that possibly could explain the difference in cell susceptibility to an external addition of $H_2O_2$, with and without a simultaneous addition of $NO_2^-$, and ultimately, quantify the effect in terms of the $H_2O_2$ membrane diffusion rate constant ($k_{D,1}$) and the intracellular catalase concentration ($[CATFe^{III}]_0$). Since we cannot distinguish a cancer cell from a normal cell solely by their $H_2O_2$ membrane diffusion rate constant and intracellular catalase concentration, we will have to work under the notations "cancer-like cells", i.e., systems in the higher range of $H_2O_2$ membrane diffusion rate constant and the lower range of catalase concentration, and "normal-like cells", i.e., systems in the lower range of $H_2O_2$ membrane diffusion rate constant and the higher range of catalase concentration. We investigate different regimes of the supplied extracellular $H_2O_2$- and $NO_2^-$ concentrations according to experimental observations of the regimes of selective/non-selective and synergistic/non-synergistic anti-cancer effect of PTL [25, 26]. The dependent variables that we investigate are:

1. **The temporal maximum of $[H_2O_2]$ and $[ONOO^-]$ in the IC.** As opposed to the steady-state value of the intracellular $H_2O_2$ concentration, the temporal maximum can be expected to be dependent on both $k_{D,1}$ and $[CATFe^{III}]_0$. These dependent variables may be related to the maximal intracellular oxidative power of the extracellularly added $H_2O_2$ (and $NO_2^-$) and thus, it would be of interest to study whether a certain extracellularly added concentration of $H_2O_2$ (and $NO_2^-$) would result in a higher oxidative power in a more cancer-like cell than in a more normal-like cell.

2. **The system response time, i.e., the time out of equilibrium, with respect to $[H_2O_2]$ in the IC.** In order to achieve tumor progression, it is essential for cancer cells to optimize their RONS concentration and maintain the RONS equilibrium. For our mathematical model, this is translated into the question: does a more cancer-like cell have a longer response time compared to a more normal-like cell?

3. **The "load" of intracellular $H_2O_2$ and $ONOO^-$, i.e., the time integral of $[H_2O_2]$ and $[ONOO^-]$ in the IC.** As the temporal maximum of $[H_2O_2]$ and $[ONOO^-]$ cannot capture any information about the total "load" of $H_2O_2$ and $ONOO^-$, i.e., how much the intracellular $[H_2O_2]$ and $[ONOO^-]$ is increased over a period of time, it could be of interest to study such a dependent variable as a complement. The load can be seen as a measure that combines the temporal maximum concentration and the system response time. Another possible way to define the load of intracellular $H_2O_2$ would be to only consider the concentration of $H_2O_2$ over a "baseline". Here, the steady-state intracellular $[H_2O_2]$, before the perturbation of an addition of extracellular $H_2O_2$ and at the upper limit of $[CATFe^{III}]_0$, is used as the baseline.

4. **The inverse of the average and maximal rate of extracellular $H_2O_2$ consumption.** Since the cell susceptibility of CAP and PTL has been found to be inversely proportional to the (extracellular) consumption rate of $H_2O_2$, it is of interest to explore a dependent variable quantifying the system susceptibility in terms of the $H_2O_2$ consumption. We investigate two such candidates where one is defined in terms of the inverse of the average $H_2O_2$ consumption rate, and the other one in terms of the inverse of the maximal $H_2O_2$ consumption rate.

For all proposed dependent variables, we will analyze the dependence on $k_{D,1}$ and $[CATFe^{III}]_0$ and whether a more cancer-like cell is associated with a higher "response" than a more normal-like cell. Our main research question is thus: Can the difference in cell susceptibility towards PTL be understood, and even quantified, by one of these dependent variables?

To the best of our knowledge, this is the first study of its kind, and our aim is to take some initial steps in the direction of an increased understanding of the mechanisms underlying the selective and synergistic anti-cancer effect of PTL, and ultimately, be able to predict the response of different cells.

## 3. Results

As introduced in section 2.2, in order to try to understand the combined role of the $H_2O_2$ membrane diffusion rate constant and the intracellular catalase concentration in determining the susceptibility of cells towards exogenous $H_2O_2$, we have to go beyond the steady-state value of the intracellular $H_2O_2$ concentration [13] (as well as latency [14]) and examine dependent variables that take the system's temporal response of a $[H_2O_2]$ perturbation in the EC into account. To be able to present the results in a more compact manner, the variables not yet introduced but of importance, and their denotations, are presented in Table 1. Details about the independent and dependent variables can be found in section 5, where the mathematical model is presented. Likewise, details about the numerical calculations, such as the values of the independent variables and parameters used in the model, can be found in section 6.

**Table 1.** Denotations of variables used in the results analysis.

| Variable | Meaning |
|---|---|
| $[H_2O_2]_0^{EC}$ | Initial $[H_2O_2]$ in the EC |
| $[NO_2^-]_0^{EC}$ | Initial $[NO_2^-]$ in the EC |
| $[H_2O_2]^{IC}$ | $[H_2O_2]$ in the IC |
| $c_{1,\max}$ | Temporal maximum of $[H_2O_2]$ in the IC |

For the analysis and interpretation of the results, we mainly consider three important features of the dependent variable of interest:

- Does it account for selectivity with respect to different regimes of $[H_2O_2]_0^{EC}$?
- Does it account for a synergistic effect when $NO_2^-$ is added to the system?
- Does it represent a feasible measure to quantify the susceptibility to exogenous $H_2O_2$ of a cell system in terms of $k_{D,1}$ and $[CATFe^{III}]_0$?

To qualify as a "measure", i.e., as a quantification of the susceptibility in terms of $k_{D,1}$ and $[CATFe^{III}]_0$, the dependent variable should be associated with a higher value for cells with a higher susceptibility and a lower value for cells with a lower susceptibility. Thus, in accordance with experimental observations, a feasible measure should result in a higher value for more cancer-like cells than for more normal-like cells, at least in the expected regime of selectivity (that is, for $[H_2O_2]_0^{EC}$ in the $\mu M$-range [26]). However, it should be noted that in this study, we do not follow strict mathematical criteria for a function to be categorized as a measure.

### 3.1 The temporal maximum of the intracellular hydrogen peroxide concentration: A possible measure of the cell susceptibility to exogenous hydrogen peroxide

Our calculation results suggest that the temporal maximum of $[H_2O_2]^{IC}$, i.e., $c_{1,\max}$, is the dependent variable of major interest in terms of our requirements. Therefore, we focus our analysis on this variable. The results of the other dependent variables are presented in Appendix A.

Figure 2 shows $c_{1,\max}$ as a function of $k_{D,1}$ and $[CATFe^{III}]_0$, for $[H_2O_2]_0^{EC} = 1\ \mu M$, with and without $NO_2^-$. The same results, but for $[H_2O_2]_0^{EC} = 1\ mM$, are shown in Figure 3.

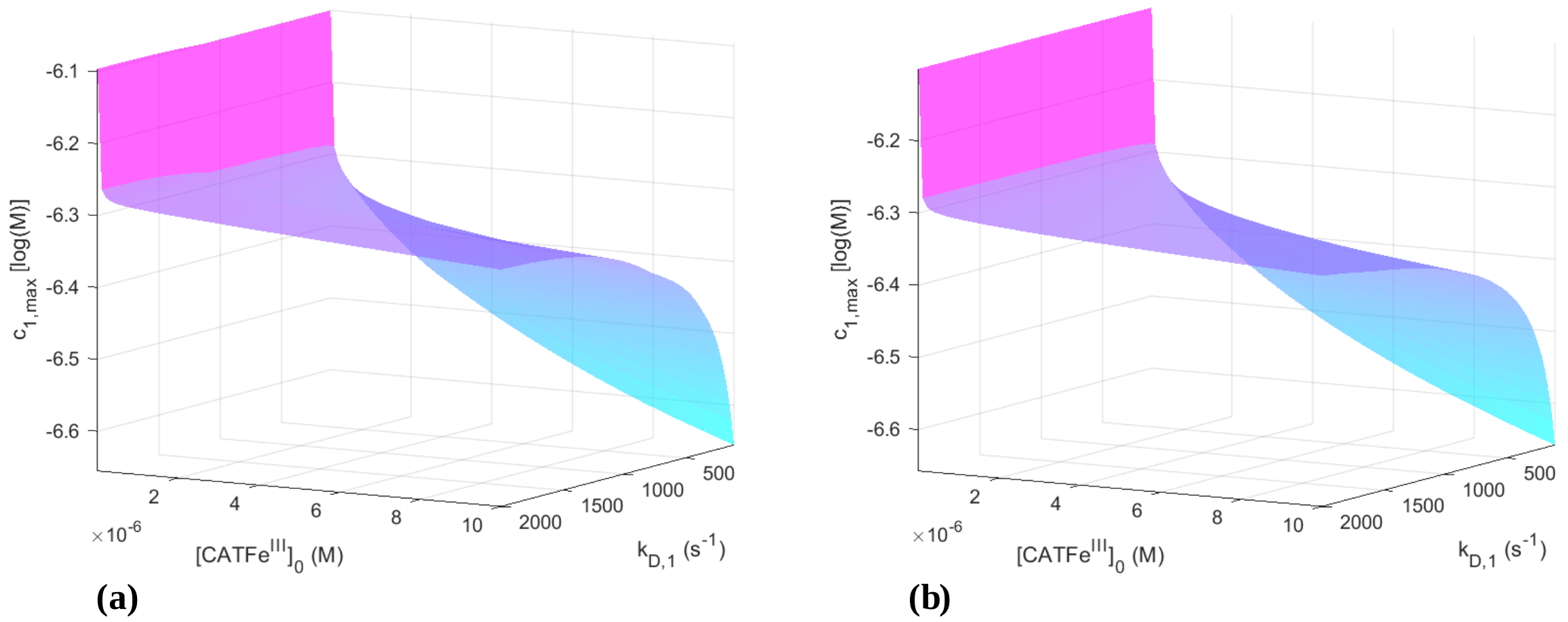

**Figure 2.** The dependent variable $c_{1,\max}$ (i.e., the temporal maximum of $[H_2O_2]$ in the IC) as a function of $k_{D,1}$ and $[CATFe^{III}]_0$ when $[H_2O_2]_0^{EC} = 1\ \mu M$. $[NO_2^-]_0^{EC} = 0\ M$ (a) and $[NO_2^-]_0^{EC} = 1\ mM$ (b).

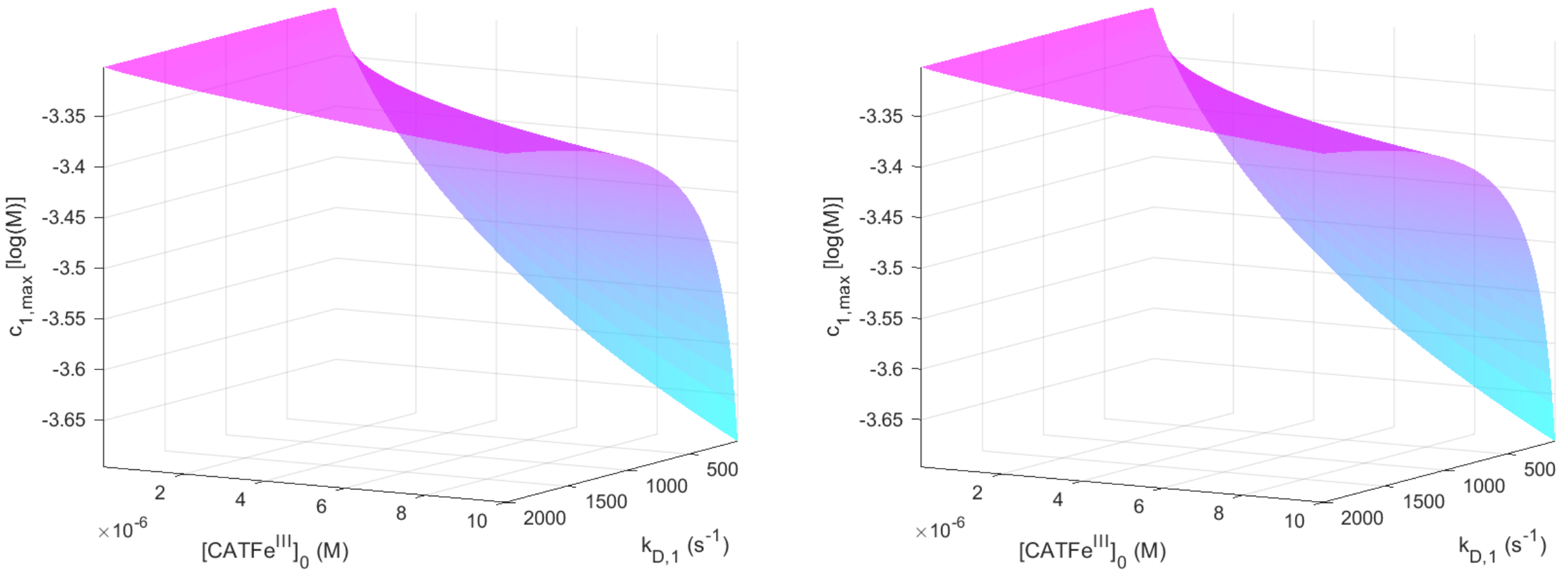

(a) (b)

**Figure 3.** The dependent variable $c_{1,\max}$ (i.e., the temporal maximum of $[H_2O_2]$ in the IC) as a function of $k_{D,1}$ and $[CATFe^{III}]_0$ when $[H_2O_2]_0^{EC} = 1\ mM$. $[NO_2^-]_0^{EC} = 0\ M$ (a) and $[NO_2^-]_0^{EC} = 1\ mM$ (b).

When comparing the result for the different $[H_2O_2]_0^{EC}$-regimes for $[NO_2^-]_0^{EC} = 0\ M$ (see Figures 2.a and 3.a), we see that $c_{1,\max}$ is also in different concentration regimes, which is logical. Indeed, for $[H_2O_2]_0^{EC} = 1\ mM$, $c_{1,\max} \gtrsim 10^{-4}\ M$, whereas for $[H_2O_2]_0^{EC} = 1\ \mu M$, $c_{1,\max} < 10^{-6}\ M$. Thus, by assuming that there exists a threshold value $c_{1,\max} > 10^{-6}\ M$ for which all types of cells undergo cell death, selectivity could be accounted for. However, there is no obvious synergetic effect; when comparing Figure 2.a and b, $c_{1,\max}$ is almost identical. Thus, the addition of $NO_2^-$ does not change $c_{1,\max}$ significantly. The same is true for Figures 3.a and b.

For $[H_2O_2]_0^{EC} = 1\ \mu M$, $c_{1,\max}$ shows an increased $k_{D,1}$-dependence with increasing $[CATFe^{III}]_0$. The lowest value of $c_{1,\max}$ is for the lowest values of $k_{D,1}$ and highest values of $[CATFe^{III}]_0$, as would be expected for a dependent variable that would qualify as a measure of the cell susceptibility in terms of $k_{D,1}$ and $[CATFe^{III}]_0$ . In addition, the highest value of $c_{1,\max}$ is associated with the lowest value of $[CATFe^{III}]_0$. However, in this regime, the dependence on $k_{D,1}$ is insignificant. Here, there is on the contrary a significant $[CATFe^{III}]_0$-dependence and by changing the scale on the $[CATFe^{III}]_0$ -axis to a log-scale (see Figure 4), we see that there are two distinct regimes with a clear shift from one regime to another at about $[CATFe^{III}]_0 \sim 10^{-7}\ M$. The regimes of $k_{D,1}$ and $[CATFe^{III}]_0$ with the most profound difference between the value of $c_{1,\max}$ is between cells with $[CATFe^{III}]_0 < 10^{-7}\ M$ (for all $k_{D,1}$) and cells with the lowest possible $k_{D,1}$ and highest possible $[CATFe^{III}]_0$. Thus, $c_{1,\max}$, i.e., the temporal maximum of $[H_2O_2]^{IC}$, is associated with a higher value for cancer-like cells than for normal-like cells. Indeed, $c_{1,\max}$ is about four times greater for the most susceptible cells compared to the most resistant cells.

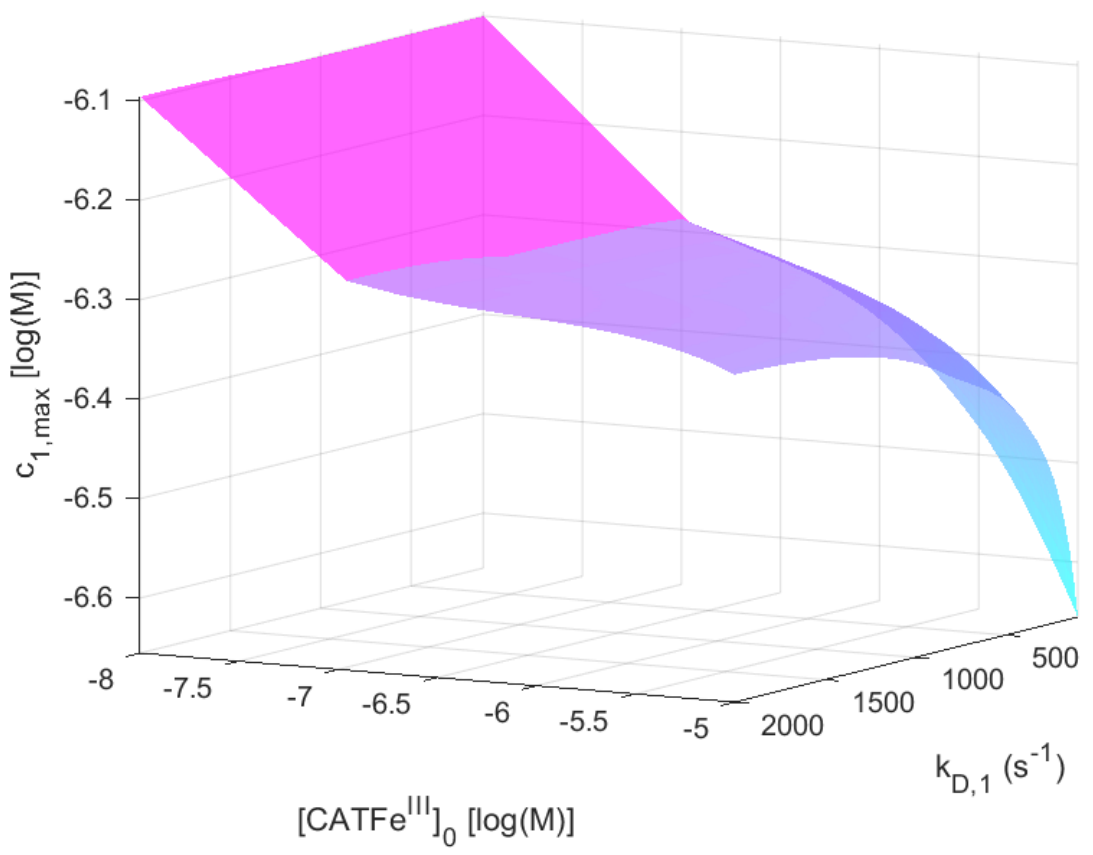

**Figure 4.** The dependent variable $c_{1,\max}$ (i.e., the temporal maximum of $[H_2O_2]$ in the IC) as a function of $k_{D,1}$ and $\log([CATFe^{III}]_0)$ when $[H_2O_2]_0^{EC} = 1\ \mu M$ and $[NO_2^-]_0^{EC} = 0\ M$.

In summary, $c_{1,\max}$ does capture the dependence of $k_{D,1}$ and $[CATFe^{III}]_0$ in a manner that is consistent with experimental observations and could thus represent a feasible measure to quantify the susceptibility of different cells in terms of their $H_2O_2$ membrane diffusion rate constant and intracellular catalase concentration. However, in our model it cannot yet account for the synergistic effect when $NO_2^-$ is added.

To the best of our knowledge, there are not yet any experimental results to support our findings. We hope that our theoretical work will inspire future experimental studies. In the next section (section 3.2), we discuss possible opportunities to experimentally quantify $c_{1,\max}$.

### 3.2 Physical interpretation and the use of the temporal maximum of the intracellular hydrogen peroxide concentration as an experimental probe

Our model, with all the equations, is explained in detail in section 5. Here we use the equations to better understand how we can use $c_{1,\max}$ as a measure to quantify the response of different cells. In order to analyze and write the equations in a more compact manner, we first introduce some short notations, as well as some new notations, see Table 2.

**Table 2.** Denotations of variables used in the results analysis.

| Variable | Meaning |
|---|---|
| $c_1^{IC}$ | $[H_2O_2]^{IC}$ |
| $c_1^{EC}$ | $[H_2O_2]^{EC}$ |
| $c_2$ | $[CATFe^{III}]$ |
| $c_3$ | $[CATFe^{IV}O^{\bullet +}]$ |
| $k_P$ | Rate of mitochondrial $H_2O_2$ production |
| $k_1$ | Rate of $H_2O_2$ consumption by $[CATFe^{III}]$ |
| $k_2$ | Rate of $H_2O_2$ consumption by $[CATFe^{IV}O^{\bullet +}]$ |

The temporal maximum of $c_1^{IC}$ (i.e., $c_{1,\max}$) occurs when the production and consumption of intracellular $H_2O_2$ are equal and $c_1^{EC} - c_1^{IC} \geq 0$ (as opposed to the steady-state value of $[H_2O_2]^{IC}$ which is goverened by the same rate equation but for which $c_1^{EC} - c_1^{IC} = 0$). If we exclude in Equation 10 (see section 5.2.2.) the term representing the formation of $ONOOH$ from $H_2O_2$ and $NO_2^-$ (since it is much smaller than the other terms), we have

$$-k_1 c_{1,\max} c_2 - k_2 c_{1,\max} c_3 + k_{D,1}\left(c_1^{EC} - c_{1,\max}\right) + k_P = 0.$$

Thus,

$$k_{D,1}\left(c_1^{EC} - c_{1,\max}\right) + k_P = k_1 c_{1,\max} c_2 + k c_{1,\max} c_3.$$

Here, we can furthermore use the constrain

$$c_3 = [CATFe^{III}]_0 - [CATFe^{III}],$$

since the total catalase concentration will be constant. By noting that $k_1 \sim k_2 = k$ (see section 6.2.1.), we can use the approximate expression

$$k_{D,1}\left(c_1^{EC} - c_{1,\max}\right) + k_P = k c_{1,\max} [CATFe^{III}]_0.$$

From our numerical calculations, we know that for low $[CATFe^{III}]_0$, $c_{1,\max}$ is independent on $k_{D,1}$, whereas for high $[CATFe^{III}]_0$, $c_{1,\max}$ is highly dependent on $k_{D,1}$. Furthermore (for $[CATFe^{III}]_0 > 10^{-7}\ M$), for low $k_{D,1}$, $c_{1,\max}$ is highly dependent on $[CATFe^{III}]_0$, whereas for high $k_{D,1}$, $c_{1,\max}$ is independent on $[CATFe^{III}]_0$. The question is whether this behavior can be understood?

In the analysis, we first note that the implicit importance of $[CATFe^{III}]_0$ and $k_{D,1}$ in determining the value of $c_1^{EC}$ at the time of $c_{1,\max}$, and thus $c_{1,\max}$, is hidden. The dependence on $[CATFe^{III}]_0$ originates from the fact that in our model, $c_{1,0}^{IC}$ is determined by $[CATFe^{III}]_0$ . Equation 3 (in section 5.2.1) can be approximated as:

$$\frac{dc_1^{EC}}{dt} = -k_{D,1}(c_1^{EC} - c_1^{IC}).$$

Hence, the initial rate, or driving force, of $H_2O_2$-consumption in the EC will crucially depend on $c_{1,0}^{IC}$, and thus, $[CATFe^{III}]_0$. In fact, $c_{1,0}^{IC} \sim 10^{-7}\ M$ for $[CATFe^{III}]_0 \sim 10^{-8}\ M$, whereas $c_{1,0}^{IC} \sim 10^{-10}\ M$ for $[CATFe^{III}]_0 \sim 10^{-5}\ M$ (see Equation (18) and Tables 5 and 8, section 6). It means that the initial driving force is about ten times higher in the latter case compared to the former. This could explain why $c_{1,\max}$ is seemingly independent on $k_{D,1}$ at low values of $[CATFe^{III}]_0$; if $k_{D,1}\left(c_1^{EC} - c_{1,\max}\right) \ll k_P$ for all values of $k_{D,1}, k_P$ will be the dominant factor of the build up of $H_2O_2$ in the IC. For higher values of $[CATFe^{III}]_0$, it seems like somewhen at $[CATFe^{III}]_0 > 10^{-7}\ M$, $c_{1,\max}$ becomes increasingly dependent on $k_{D,1}$. It is thus reasonable to believe that the term $k_{D,1}\left(c_1^{EC} - c_{1,\max}\right)$ is becoming increasingly dominant and that the larger the value of $[CATFe^{III}]_0$, the larger the value of $\left(c_1^{EC} - c_{1,\max}\right)$. For a fixed value of $[CATFe^{III}]_0$, $c_{1,\max}$ will thus increase with increasing values of $k_{D,1}$.

In summary, this means that the lower the value of $[CATFe^{III}]_0$, the less important is the value of $k_{D,1}$, and the other way around. Thus, the susceptibility (towards exogenous $H_2O_2$) of cancer-like cells is not much influenced by the $H_2O_2$ membrane diffusion rate constant and this is due to their much higher level of intracellular $H_2O_2$ prior to the perturbation by addition of exogenous $H_2O_2$. Normal-like cells, on the other hand, are more sensitive to the value of the $H_2O_2$ membrane diffusion rate constant, since the difference in concentration between the intracellular and extracellular $H_2O_2$ will be much larger.

Another aspect of $c_{1,\max}$ is whether it could provide an opportunity to extract information about different cell lines in terms of their $H_2O_2$ membrane diffusion rate constant and intracellular catalase concentration. By measuring $c_{1,\max}$ and the corresponding $c_1^{EC}$ for different $c_{1,0}^{EC}$ it could be possible to roughly quantify $k_{D,1}$ and $[CATFe^{III}]_0$. There are many experimental techniques for detection and quantification of the $H_2O_2$ concentration *in vitro* and *in vivo*. The intracellular $H_2O_2$ concentration has e.g. been detected and measured by a chemoselective fluorescent naphthylimide peroxide probe [70], by a genetically encoded red fluorescent sensor [71], and by fluorescent reporter proteins [72]. Thus, even if $c_{1,\max}$ does not represent a feasible measure of the cell susceptibility in terms of $k_{D,1}$ and $[CATFe^{III}]_0$, it could still possibly be used to gain more knowledge about the correlation between $k_{D,1}$ and $[CATFe^{III}]_0$ and cell susceptibility towards exogenous $H_2O_2$ and PTLs.

## 4. Discussion

In this study, we use a theoretical approach to increase the knowledge about possible underlying causes of the anti-cancer effect of PTL. Although the model is fairly simple, it does include the major pathways for species production and consumption relevant for such a cell system. It also puts emphasis on two important features (i.e., the $H_2O_2$ membrane diffusion rate constant and the intracellular catalase concentration), possibly explaining the different cell responses and cell susceptibility towards PTL, when comparing normal cells to cancer cells, but also when comparing resistant vs sensitive cancer cells. Nevertheless, it is important to keep in mind that in our model, different cells are only defined in terms of these two features, which are independent variables in our analysis, whereas in reality there are countless of other features characteristic for different types of cells, that could play an important role in the context of the anti-cancer effect of PTL. Here, we merely analyze the immediate cell response determined by the scavenging system active at high concentrations of $H_2O_2$. However, we do believe that our results contribute to a better understanding of some mechanisms probably underlying the anti-cancer effect of PTL. It brings novelty to the field of plasma oncology, and more broadly, to the field of redox biology, by using a theoretical approach and by proposing new ways to quantify the selective and synergistic anti-cancer effect of PTL in terms of inherent factors of cells. Here, we discuss each of our main findings and their potential implications. We also highlight what we believe are the most important limitations of the model.

As opposed to the steady-state intracellular concentration of $H_2O_2$, which has been evaluated in previous studies [13], our results suggest that the temporal maximal concentration of intracellular $H_2O_2$ could be a measure feasible to quantify the cell susceptibility towards exogenous $H_2O_2$ in terms of the $H_2O_2$ membrane diffusion rate constant and the intracellular catalase concentration. This result furthermore enables us to speculate whether the mode of action of $H_2O_2$ is as a signaling molecule rather than as a toxic substance causing necrosis. It is known that the intracellular concentration of a signaling molecule rises and falls within a short period. Indeed, whether a signaling molecule is effective or not, is determined by how rapidly it is produced, how rapidly it is removed, and the concentration it must reach to alter the activity of its target effector. Of particular relevance in our context is that several reports have demonstrated that the rate of $H_2O_2$ generation and its concentration as a function of time play a key role in determining target cell damage or destruction [73-75]. RONS are regulators of signaling pathways, such as the extracellular signal-regulated kinase (ERK) mitogen activated protein kinase (MAPK) pathway, which is important for cell proliferation, and a number of studies have demonstrated the ability of exogenous oxidants to activate the ERK MAPK pathway [76-80]. As in the general case, the duration and intensity of the ERK MAPK signal determine the outcome of the cellular response; there is a connection between the levels of ROS in a cell and the levels of MAPK signaling. Especially, MAPKs are activated in response to $H_2O_2$ [81-83].

Based on our modeling results (presented in the Appendix A; i.e., figures A1 and A9), we do not think that the formation of $ONOO^-$ itself plays a major role in the explanation of the synergistic effect of $H_2O_2$ and $NO_2^-$. This is because although the overall intracellular concentration of $ONOO^-$ is increased with about one order of magnitude when $NO_2^-$ is added to the system, the dependence on the $H_2O_2$ membrane diffusion rate constant is such that cells with a higher value of the $H_2O_2$ membrane diffusion rate constant, i.e., cancer-like cells, are associated with a lower maximal intracellular $ONOO^-$ concentration than more normal-like cells (i.e., cells with a lower value of the $H_2O_2$ membrane diffusion rate constant). In addition, the load of $ONOO^-$ is independent of the $H_2O_2$ membrane diffusion rate constant. However, an important aspect to keep in mind regarding our results for $ONOO^-$ and the choice to include $CO_2$-catalyzed consumption of $ONOO^-$ in our model, is that $CO_2$ redirects much of the $ONOO^-$ produced *in vivo* towards radical mechanisms [65]. Indeed, many of the reactions of $ONOO^-$ *in vivo* are more likely to be mediated by reactive intermediates derived from the reaction of $ONOO^-$ with $CO_2$ than by $ONOO^-$ itself [84, 85]. Thus, if the production of such reactive intermediates were to be monitored instead of $ONOO^-$, our results might be different. In this context, especially the formation of $CO_3^{\bullet-}$ should be considered; a fraction (about 30%) of the formed $ONOOCOO^-$ will produce cage-escaped $^{\bullet}NO_2$ and $CO_3^{\bullet-}$ radicals according to [86-88]

$$ONOOCOO^- \rightleftharpoons CO_3^{\bullet-} + {}^{\bullet}NO_2,$$

where $k = 1.9 \times 10^9\ s^{-1}$ and $k' = 5 \times 10^8\ M^{-1}s^{-1}$ [89]. A possibly important target in the context of our study, is catalase; catalase is so far the best known protein target for $CO_3^{\bullet-}$ and the rate constant of the reaction of bovine liver catalase with $CO_3^{\bullet-}$is $(3.7 \pm 0.4) \times 10^9\ M^{-1}s^{-1}$ at $pH = 8.4$ [90]. Since the temporal maximum of intracellular $[H_2O_2]$, i.e., $c_{1,\max}$, is inversely dependent on the catalase concentration, i.e., $[CATFe^{III}]_0$, with an increasingly steeper incline for lower catalase concentrations in the regime $10^{-8} \leq [CATFe^{III}]_0 \leq 10^{-7}\ M$ (see Figure 2), cancer-like cells would be more vulnerable to a decrease in the catalase concentration than normal-like cells, which are associated with higher values of $[CATFe^{III}]_0$. Thus, including these reaction pathways may possibly also make the dependent variable $c_{1,\max}$ able to account for the synergetic effect of $NO_2^-$. Such an extension of our model was out of the scope for this study, but would be highly interesting in a future model development.

In experiments, the consumption rate of extracellular $H_2O_2$ has been found to inversely correlate with the susceptibility of cancer cells towards exogenous $H_2O_2$ [32]. Thus, cancer cell lines with a high consumption rate were less susceptible. Our results cannot yet account for this correlation; when cells are defined in terms of their $H_2O_2$ membrane diffusion rate constant and their intracellular catalase concentration, susceptibility in terms of the inverse of the extracellular $H_2O_2$ consumption rate is not consistent with the experimental observations of cancer cells having a higher $H_2O_2$ membrane diffusion rate constant and a lower catalase concentration (see section 2.1). The fact that our model does not reproduce these patterns leaves an open question of how to construct a dependent variable in terms of the inverse of the extracellular $H_2O_2$ consumption rate such that it corresponds to the experimental correlation.

The fact that our mathematical model, as well as our criteria for a dependent variable to represent a feasible measure of the cell susceptibility, do not select the system response time as a good candidate, does not necessarily indicate that this variable in general cannot capture cell susceptibility towards exogenous $H_2O_2$. Indeed, in our definition of this dependent variable we assume a tolerance of a 10% increase of the intracellular steady-state $H_2O_2$ concentration, and a different assumption of the tolerance might give a different result.

Except for the limitations of the model already mentioned in this discussion, some other model assumptions could hamper a realistic representation of a cell system in interaction with PTL. One such limitation is that in our model, the rate of mitochondrial $H_2O_2$ production is constant. Although it can be argued that this assumption is a valid starting point, in a model development it could be important to modify this aspect to represent a cancer cell in a more realistic manner. Indeed, it has been shown that in some cancer cells, the mitochondrial respiration is decreased (in favor of aerobic glycolysis) and moreover this shift seems to be a dynamic process (see e.g. ref. [91] and references therein). We believe that future models could benefit from trying to take such variation of the rate of mitochondrial $H_2O_2$ production into account, but this was out of the scope for this study.

Another aspect to take into account in a more realistic model is the fact that the $H_2O_2$ membrane diffusion rate constant is not a static but dynamic property. In e.g. ref. [92] it was shown that cellular stress conditions reversibly inhibit the diffusion of $H_2O_2$ (and $H_2O$) of aquaporin 8. Thus, a more complex model taking the implicit time-dependence of the $H_2O_2$ membrane diffusion rate (caused by the increased intracellular $H_2O_2$ concentration after the addition of exogenous $H_2O_2$) could potentially produce results different from our model.

A third aspect to be aware of is that in our model, we assume that the addition of PTL does not affect the membrane diffusion rate constants. However, a number of studies have reported an enhanced cell membrane permeability (and thus, increased membrane diffusion rate constants) after CAP/PTL treatment [93-95]. For the aim and approach of our study, where the membrane diffusion rate constant of the key species $H_2O_2$ is varied within a range of possible values, we believe that our assumption is a valid starting point. Nevertheless, for future model extensions and developments, this aspect might be important to take into account.

Finally, it should be mentioned that the rate equations used to model the system are derived from information (collected from the literature) about rate constants and reaction orders for each reaction as they appear in experiments. Most likely, the experimental conditions will deviate from the conditions of cells treated with PTL, which will affect the accuracy of the results produced by the model. However, for the purpose of our study, we believe that parameter values of the correct order of magnitude are sufficient at this stage.

# 5. Mathematical model

Mathematical models of biological reaction networks, such as the system considered in this study, can generally be divided into two categories: predictive and descriptive models. Since the experimental studies on which we build our model on are primarily *in vitro* studies, we construct a predictive model in this work. This means that we put together the information about each of the involved reactions (reaction orders, rate constants, etc.) as they appear in experiments. From there, the result for a certain set of initial conditions is generated by solving the time-dependent equations of motion, representing the time evolution of the system.

In this section, we systematically present the species and reactions in the system considered (section 5.1) and how the system time evolution is modeled (section 5.2), and we explicitly define the dependent variables that are analyzed (section 5.3).

## 5.1. Species and reactions in the system

The involved species of interest are $H_2O_2$, $ONOO^-$, $ONOOH$, $NO_2^-$, $CO_2$, $H^+$, $CATFe^{III}$ and $CATFe^{IV}O^{\bullet +}$ (see Figure 1 in section 2). The following reactions and interactions of the species $H_2O_2$, $NO_2^-$ and (native) catalase ($CATFe^{III}$) in the system are taken into account.

*Decomposition of $H_2O_2$ by catalase:*

$$CATFe^{III} + H_2O_2 \xrightarrow{k_1} CATFe^{IV}O^{\bullet +} + H_2O,$$

$$CATFe^{IV}O^{\bullet +} + H_2O_2 \xrightarrow{k_2} CATFe^{III} + O_2 + H_2O.$$

*Generation of $ONOO^-$ through reaction between $H_2O_2$ and $NO_2^-$:*

$$NO_2^- + H_2O_2 + H^+ \xrightarrow{k_3} ONOOH + H_2O,$$

where the equilibrium between $ONOOH$ and $ONOO^-$ is described by:

$$ONOOH \underset{k_{-4}}{\overset{k_4}{\rightleftharpoons}} ONOO^- + H^+.$$

*Decomposition of $ONOO^-$ by catalase:*

$$2ONOO^- \underset{CATFe^{III}}{\overset{k_5}{\rightarrow}} O_2 + 2NO_2^-.$$

*$CO_2$-catalyzed consumption of $ONOO^-$:*

$$ONOO^- + CO_2 \xrightarrow{k_6} ONOOCOO^-.$$

The denotations of the time-dependent concentrations of the different species are shown in Table 3.

**Table 3.** Denotations of the time-dependent concentrations in the system.

| Species | Denotation |
|---|---|
| $[H_2O_2]$ | $c_1$ |
| $[CATFe^{III}]$ | $c_2$ |
| $[CATFe^{IV}O^{\bullet +}]$ | $c_3$ |
| $[ONOO^-]$ | $c_4$ |
| $[NO_2^-]$ | $c_5$ |
| $[H^+]$ | $c_6$ |
| $[ONOOH]$ | $c_7$ |
| $[CO_2]$ | $c_8$ |

### 5.2. Modeling the system

The mathematical model considers the kinetics of the reactions in the system composed of two subsystems (EC and IC), see Figure 1 in section 2, as well as diffusion of certain species between the two subsystems. The equation governing the kinetics of each species $i$ is given by the sum of the reaction rates (describing the rate of production and consumption of species $i$),

$$\frac{dc_i}{dt} = r_i, \tag{1}$$

and (in the case of species $1,\ 4,\ 5$ and $7$), the diffusion rate through the cell membrane, from the EC to the IC,

$$\frac{dc_i}{dt} = -k_{D,i}(c_i^{EC} - c_i^{IC}). \tag{2}$$

Equation (1) represents the resulting rate equation, derived from the rate constants and reaction orders for each reaction as they appear in experiments. Equation (2) describes the rate of membrane diffusion of species $i$ according to Fick's law of diffusion with a linear concentration gradient over the cell membrane. Here, $k_{D,i}$ is the rate of species $i$ exchange through the membrane. We denote this as "membrane diffusion rate constant". More information about the derivation of Equation 2 can be found in Appendix B.

Explicitly, our mathematical model is used to analyze the behavior of a dependent variable $y(\bar{x})$, where $\bar{x}$ denotes the set of independent variables that are varied in the system. The independent variables in our model are:

- The $H_2O_2$ membrane diffusion rate constant through the cell membrane ($k_{D,1}$)
- The initial intracellular catalase concentration ($[CATFe^{III}]_0$)

The species $CO_2$ and $H^+$ are assumed to be present in equal initial concentrations in both the EC and the IC (thus, $c_{8,0}^{EC} = c_{8,0}^{IC}$ and $c_{6,0}^{EC} = c_{6,0}^{IC}$) . Since we do not explicitly study the kinetics of these species, we make such an assumption to reduce the complexity of the model.

Detailed information about the mathematical model is presented in the following sections.

#### *5.2.1. Mathematical model of the reaction kinetics in the extracellular compartment*

At $t = 0$, $H_2O_2$ and $NO_2^-$ in certain initial concentrations ($c_{1,0}^{EC} = [H_2O_2]_0^{EC}$ and $c_{5,0}^{EC} = [NO_2^-]_0^{EC}$) are inserted into the EC, representing treatment of the cell by PTL (as these species are the dominant RONS in PTLs), and their reactions as well as diffusion through the membrane into the IC is monitored. The reaction network and resulting set of differential equations are given below.

*Reaction network*

$$NO_2^- + H_2O_2 + H^+ \xrightarrow{k_3} ONOOH + H_2O,$$

$$ONOOH \underset{k_{-4}}{\overset{k_4}{\rightleftharpoons}} ONOO^- + H^+,$$

$$ONOO^- + CO_2 \xrightarrow{k_6} ONOOCOO^-,$$

$$H_2O_2 \xrightarrow{k_{D,1}} IC,$$

$$ONOO^- \xrightarrow{k_{D,4}} IC,$$

$$NO_2^- \xrightarrow{k_{D,5}} IC,$$

$$ONOOH \xrightarrow{k_{D,7}} IC.$$

*Differential equations*

$$\frac{dc_1^{EC}}{dt} = -k_3 c_1^{EC} c_5^{EC} c_6^{EC} - k_{D,1}(c_1^{EC} - c_1^{IC}), \quad (3)$$

$$\frac{dc_4^{EC}}{dt} = k_4 c_7^{EC} - k_{-4} c_4^{EC} c_6^{EC} - k_6 c_4^{EC} c_8^{EC} - k_{D,4}(c_4^{EC} - c_4^{IC}), \quad (4)$$

$$\frac{dc_5^{EC}}{dt} = -k_3 c_1^{EC} c_5^{EC} c_6^{EC} - k_{D,5}(c_5^{EC} - c_5^{IC}), \quad (5)$$

$$\frac{dc_6^{EC}}{dt} = -k_3 c_1^{EC} c_5^{EC} c_6^{EC} + k_4 c_7^{EC} - k_{-4} c_4^{EC} c_6^{EC}, \quad (6)$$

$$\frac{dc_7^{EC}}{dt} = k_3 c_1^{EC} c_5^{EC} c_6^{EC} - k_4 c_7^{EC} + k_{-4} c_4^{EC} c_6^{EC} - k_{D,7}(c_7^{EC} - c_7^{IC}), \quad (7)$$

$$\frac{dc_8^{EC}}{dt} = -k_6 c_4^{EC} c_8^{EC} \;. \quad (8)$$

*5.2.2. Mathematical model of the reaction kinetics in the intracellular compartment*

At $t = 0$, the concentration of $H_2O_2$ is at a certain steady-state value ($c_{1,0}^{IC} = [H_2O_2]_0^{IC}$), because it is continuously produced by the mitochondria at the rate

$$\frac{dc_1^{IC}}{dt} = k_P, \quad (9)$$

and decomposed by catalase, which exists in the IC, and is modeled as free in the solution. (More information can be found in section 6.2.3). It is assumed that at $t = 0$, the total amount of catalase exists as $CATFe^{III}$, i.e., $c_{3,0}^{IC} = 0$. The reaction network and resulting set of differential equations are given below.

*Reaction network*

$$CATFe^{III} + H_2O_2 \xrightarrow{k_1} CATFe^{IV}O^{\bullet+} + H_2O,$$

$$CATFe^{IV}O^{\bullet+} + H_2O_2 \xrightarrow{k_2} CATFe^{III} + O_2 + H_2O,$$

$$NO_2^- + H_2O_2 + H^+ \xrightarrow{k_3} ONOOH + H_2O,$$

$$ONOOH \underset{k_{-4}}{\overset{k_4}{\rightleftharpoons}} ONOO^- + H^+,$$

$$2ONOO^- \xrightarrow{k_5} O_2 + 2NO_2^-,$$

$$ONOO^- + CO_2 \xrightarrow{k_6} ONOOCOO^-,$$

$$EC \xrightarrow{k_{D,1}} H_2O_2,$$

$$EC \xrightarrow{k_{D,4}} ONOO^-,$$

$$EC \xrightarrow{k_{D,5}} NO_2^-,$$

$$EC \xrightarrow{k_{D,7}} ONOOH.$$

*Differential equations*

$$\frac{dc_1^{IC}}{dt} = -k_1 c_1^{IC} c_2 - k_2 c_1^{IC} c_3 - k_3 c_1^{IC} c_5^{IC} c_6^{IC} + k_{D,1}(c_1^{EC} - c_1^{IC}) + k_P, \quad (10)$$

$$\frac{dc_2}{dt} = -k_1 c_1^{IC} c_2 + k_2 c_1^{IC} c_3 , \quad (11)$$

$$\frac{dc_3}{dt} = k_1 c_1^{IC} c_2 - k_2 c_1^{IC} c_3, \quad (12)$$

$$\frac{dc_4^{IC}}{dt} = k_4 c_7^{IC} - k_{-4} c_4^{IC} c_6^{IC} - k_5 c_2 c_4^{IC} - k_6 c_4^{IC} c_8^{IC} + k_{D,4}(c_4^{EC} - c_4^{IC}) , \quad (13)$$

$$\frac{dc_5^{IC}}{dt} = -k_3 c_1^{IC} c_5^{IC} c_6^{IC} + k_5 c_2 c_4^{IC} + k_{D,5}(c_5^{EC} - c_5^{IC}) , \quad (14)$$

$$\frac{dc_6^{IC}}{dt} = -k_3 c_1^{IC} c_5^{IC} c_6^{IC} + k_4 c_7^{IC} - k_{-4} c_4^{IC} c_6^{IC}, \quad (15)$$

$$\frac{dc_7^{IC}}{dt} = k_3 c_1^{IC} c_5^{IC} c_6^{IC} - k_4 c_7^{IC} + k_{-4} c_4^{IC} c_6^{IC} + k_{D,7}(c_7^{EC} - c_7^{IC}), \quad (16)$$

$$\frac{{dc_8}^{IC}}{dt} = -k_6 c_4^{IC} c_8^{IC}. \quad (17)$$

Equations (3-8) and (10-17) are solved numerically. The details about the numerical calculations can be found in section 6.3.

**5.3. Dependent variables**

In the following sections we explicitly define the dependent variables analyzed in this study.

*5.3.1. Temporal maximum of the intracellular hydrogen peroxide and peroxynitrite concentration*

The dependent variables $c_{1,\max}$ and $c_{4,\max}$ are defined as

$$c_{1,\max}\left(k_{D,1},[CATFe^{III}]_0\right)=\max([H_2O_2]^{IC}),$$

and

$$c_{4,\max}\left(k_{D,1},[CATFe^{III}]_0\right)=\max([ONOO^-]^{IC}).$$

*5.3.2. System response time of intracellular hydrogen peroxide*

Assuming that the system has a tolerance of an increase of 10 % of the baseline $H_2O_2$ concentration (see section 3.2), the dependent variable $\tau$ can be formulated

$$\tau\left(k_{D,1},[CATFe^{III}]_0\right)=t\ni\frac{(100+10)}{100}\times[H_2O_2]_t^{IC}=[H_2O_2]_0^{IC}.$$

*5.3.3. Load of intracellular hydrogen peroxide and peroxynitrite*

The simplest way of creating a quantitative measure of the "load" of intracellular $H_2O_2$ and $ONOO^-$ is to use the time-integral over the whole time regime ($0\le t\le t_f$) as the dependent variable, i.e.,

$$l_1\left(k_{D,1},[CATFe^{III}]_0\right)=\int_0^{t_f}[H_2O_2]^{IC},$$

and

$$l_4\left(k_{D,1},[CATFe^{III}]_0\right)=\int_0^{t_f}[ONOO^-]^{IC}.$$

For the "load" over the baseline concentration of intracellular $H_2O_2$; if we denote this baseline constant $[H_2O_2]_{BS}^{IC}$, the dependent variable is defined as

$$l_{1,BS}(k_{D,1},[CATFe^{III}]_0)=\int_0^{t_f}([H_2O_2]^{IC}-[H_2O_2]_{BS}^{IC}).$$

*5.3.4. Rate of extracellular hydrogen peroxide consumption*

Here, we first define the dependent variable $r$ as

$$r(k_{D,1},[CATFe^{III}]_0)=\frac{d[H_2O_2]^{EC}}{dt}.$$

The average extracellular consumption rate of $H_2O_2$ is then defined as

$$\overline{r}=\frac{1}{t_f}\int_0^{t_f}r\,dt,$$

and the maximal extracellular consumption rate of $H_2O_2$ as

$$r_{\max} = \max(|r|).$$

In order to create a potential measure, i.e., a dependent variable where a more cancer-like cell is associated with a higher susceptibility, we use the variables

$$\bar{s} = \frac{1}{\bar{r}},$$

and

$$s_{\max} = \frac{1}{r_{\max}},$$

in our calculations.

# 6. Numerical calculations

## 6.1 Independent variables

The two independent variables in the system are $k_{D,1}$ and $c_{2,0} = [CATFe^{III}]_0$, i.e., the diffusion rate constant of $H_2O_2$ through the cell membrane from the EC to the IC, and the initial catalase concentration in the IC. Furthermore, we use four different combinations of $c_{1,0}^{EC} = [H_2O_2]_0^{EC}$ and $c_{5,0}^{EC} = [NO_2^-]_0^{EC}$ in our calculations. The motivation and details of these variables are given in the following sections, and have also been introduced in section 2.

### *6.1.1. Membrane diffusion rate constant of hydrogen peroxide*

In ref. [96], the diffusion rate constant for $H_2O$ crossing lipid bilayers was found to be $920\ s^{-1}$. Due to the chemical similarities between $H_2O$ and $H_2O_2$, we use this value as a reference value for $k_{D,1}$ and we vary $k_{D,1}$ within the range $100 \leq k_{D,1} \leq 2000\ s^{-1}$.

### *6.1.2. Initial catalase concentration in the intracellular compartment*

The intracellular concentration of catalase is calculated from two different premises, see Appendix C. Considering the rough estimates in both approaches, it seems reasonable to use an effective catalase concentration in the range of $10^{-8} - 10^{-5}\ M$ in our calculations. As a reference value, catalase concentration in human blood cells is about $2 - 3\ \mu M$ [97, 98].

### *6.1.3. Initial hydrogen peroxide and peroxynitrite concentration in the extracellular compartment*

Several publications have shown that $H_2O_2$ and $NO_2^-$ are formed at concentrations ranging from $\mu M$ to $mM$ in plasma-treated liquids (PTLs) [99-102]. In this study, we use the initial conditions for $[H_2O_2]^{EC}$ and $[NO_2^-]^{EC}$ shown in Table 4. The different regimes of these four combinations are specified in the last column. We assume that the *selectivity* is related to the concentration of extracellular $H_2O_2$, i.e., selective cancer killing only occurs at low $H_2O_2$ concentrations (order of 1 $\mu M$ [26]), while higher $H_2O_2$ concentrations (e.g., order of 1 $mM$) kill both cancer and normal cells [25] . Based on this assumption we want to compare the dependent variables for the selective versus non-selective regime. For both regimes (selective versus non-selective), we furthermore want to investigate whether a *synergistic effect* can be found, i.e., if the values of the dependent variables are enhanced when $H_2O_2$ and $NO_2^-$ are added together [25, 26].

**Table 4.** Initial concentrations of $H_2O_2$ and $NO_2^-$ in the extracellular compartment.

| $[H_2O_2]^{EC}$ $(M)$ | $[NO_2^-]^{EC}$ $(M)$ | Regime |
|---|---|---|
| $10^{-3}$ | $10^{-3}$ | Non-selective, synergistic |
| $10^{-3}$ | 0 | Non-selective, non-synergistic |
| $10^{-6}$ | $10^{-3}$ | Selective, synergistic |
| $10^{-6}$ | 0 | Selective, non-synergistic |

## 6.2. Parameter values

### *6.2.1. Reaction rate constants*

The used rate constants are summarized in Table 5, along with the references where the data is adopted from, and some remarks about the conditions for which these values were reported.

**Table 5.** Reaction rate constants.

| Rate constant | Parameter value | Reference | Remark |
|---|---|---|---|
| $k_1$ | $1.7 \times 10^7\ M^{-1}s^{-1}$ | [103] | Mammalian catalases |
| $k_2$ | $2.6 \times 10^7\ M^{-1}s^{-1}$ | [103] | Mammalian catalases |
| $k_3$ | $1.1 \times 10^3\ M^{-2}s^{-1}$ | [37] | At $pH = 3.3$ and $T = 25$ ℃ |
| $k_4$ | $K_a k_{-4} = 10^{-pK_a} k_{-4}$ | [104, 105] | The $pK_a$-value at $T = 25$ ℃ is $6.5 - 6.8$ |
| $k_{-4}$ | $\sim 10^{10}\ M^{-1}s^{-1}$ | [106] | |
| $k_5$ | $1.7 \times 10^6\ M^{-1}s^{-1}$ | [64] | At $pH = 7.1$ and $T = 25$ ℃ |
| $k_6$ | $5.8 \times 10^4\ M^{-1}s^{-1}$ | [84] | At $T = 37$ ℃, $pH$-independent |

### *6.2.2. Membrane diffusion rate constants*

$NO_2^-$, when protonated (i.e., as $HNO_2$), is reported to diffuse easily across biological membranes [107]. When not protonated, anionic channels have been shown to be permeable to $NO_2^-$ [108]. It has furthermore been established that $ONOO^-$ is able to penetrate cell membranes [96, 109]. In ref. [96], using model phospholipid vesicular systems, it was demonstrated that $ONOO^-$ freely crosses phospholipid membranes. The diffusion rate constant for $ONOO^-$ crossing lipid bilayers was found to be $k_{D,4} = 320\ s^{-1}$. Due to the acid-base equilibrium between $ONOO^-$ and its conjugated acid $ONOOH$, this is likely an average value for $ONOO^-$ and $ONOOH$. Thus, $k_{D,7} = k_{D,4} = 320\ s^{-1}$. Since $NO_2^-$ is an anion as well as similar in size, we assume the same value, i.e. $k_{D,4} = k_{D,5}$. The used diffusion rate constants are summarized in Table 6. Note that we do not consider the potential effect of the PTL on the membrane diffusion rate constants in our model.

**Table 6.** Membrane diffusion rate constants.

| Rate constant | Parameter value | Reference | Remark |
|---|---|---|---|
| $k_{D,4}$ | $320\ s^{-1}$ | [96] | |
| $k_{D,5}$ | $320\ s^{-1}$ | | Assigned |
| $k_{D,7}$ | $320\ s^{-1}$ | [96] | |

### *6.2.3. Initial concentrations*

The used initial concentrations in the EC and IC are summarized in Tables 7 and 8, respectively.

**Table 7.** Initial concentrations of the species in the extracellular compartment.

| Species | Initial concentration ($M$) | Reference | Remark |
|---|---|---|---|
| $H_2O_2$ | Varied | | |
| $ONOO^-$ | 0 | | Assigned |
| $NO_2^-$ | Varied | | |
| $H^+$ | $10^{-7}$ | | Assigned |
| $ONOOH$ | 0 | | Assigned |
| $CO_2$ | $10^{-3}$ | [110] | |

**Table 8.** Initial concentrations of the species in the intracellular compartment.

| Species | Initial concentration ($M$) | Reference | Remark |
|---|---|---|---|
| $H_2O_2$ | Varied | | |
| $CATFe^{III}$ | Varied | | |
| $CATFe^{IV}O^{\bullet+}$ | 0 | | Assigned |
| $ONOO^-$ | 0 | | Assigned |
| $NO_2^-$ | $10^{-4}$ | [111-113] | See Appendix D |
| $H^+$ | $10^{-7}$ | | Assigned |
| $ONOOH$ | 0 | | Assigned |
| $CO_2$ | $10^{-3}$ | [110] | See Appendix D |

The initial concentration of intracellular $H_2O_2$, i.e. $c_{1,0}^{IC}$, is varied with the initial concentration of catalase in order to achieve the correct steady-state $c_{1,0}^{IC}$ for each $[CATFe^{III}]_0$. The $H_2O_2$-generation from mitochondria is in the range of 50 $\mu mol\ kg^{-1}\ min^{-1}$ [114], which corresponds to $k_P = 1 \times 10^{-7}\ Ms^{-1}$ [115].

Thus, from Equation (10), at $t = 0$ (and thus, the term $k_{D,1}(c_1^{EC} - c_1^{IC})$ vanishes),

$$\frac{dc_1^{IC}}{dt} = -k_1 c_1^{IC} c_2 - k_2 c_1^{IC} c_3 - k_3 c_1^{IC} c_5^{IC} c_6^{IC} + k_P = 0.$$

Assuming that $c_3 = 0$ at steady-state,

$$k_P = c_{1,ss}^{IC}(k_1 c_2 + k_3 c_5^{IC} c_6^{IC}) \Leftrightarrow$$
$$c_{1,ss}^{IC} = \frac{k_P}{(k_1 c_2 + k_3 c_5^{IC} c_6^{IC})}. \quad (18)$$

Hence, $c_{1,ss}^{IC} = c_{1,0}^{IC}$ in our model.

### 6.3. Software and details about the calculations

The numerical calculations are performed in MATLAB. Due to significant differences in time scales, we use the solver ode23s to solve the set of rate equations.

The simulations are performed at time-scales covering the transient of the system's response. For the calculations, we use the time intervals and time steps shown in Table 9. We start with a very short time-step in the first 10 ms, which is then enlarged by a factor 100 until 1 s, and again by a factor 100 until the final time of 100 s.

**Table 9.** Time intervals and time steps.

| Time | Value (s) | Time step | Value (s) |
|---|---|---|---|
| $t_1$ | $10^{-2}$ | $dt_1$ | $10^{-7}$ |
| $t_2$ | 1 | $dt_2$ | $10^{-5}$ |
| $t_f$ | $10^2$ | $dt_3$ | $10^{-3}$ |

We furthermore vary the independent variables according to Table 10.

**Table 10.** Minimal and maximal values, as well as number of steps, of independent variables.

| Independent variable | Minimal value | Maximal value | Number of steps |
|---|---|---|---|
| $[CATFe^{III}]$ | $10^{-8}\ M$ | $10^{-5}\ M$ | 100 |
| $k_{D,1}$ | $100\ s^{-1}$ | $2000\ s^{-1}$ | 100 |

# 7. Conclusions

With this study, we aim to gain insights about mechanisms possibly underlying the anti-cancer effect of plasma-treated liquids (PTLs). Especially, we are interested in whether cell susceptibility towards PTL can be quantified in terms of cell-specific features, how selectivity arises, and why $H_2O_2$ combined with $NO_2^-$ (as in PTL) offers a synergistic and thus enhanced anti-cancer effect as compared with $H_2O_2$ only. By developing a mathematical model describing the kinetics of the species in PTL-treated cells, we analyze four different dependent variables as a function of the $H_2O_2$ membrane diffusion rate constant and the intracellular catalase concentration. Ultimately, one or more of these dependent variables could be used to quantify selective and synergistic effects of PTLs for different types of cells. In accordance with experimental observations, cancer cells are supposed to be associated with a higher $H_2O_2$ membrane diffusion rate constant and a lower intracellular catalase concentration, as compared to normal cells, and we use this knowledge in the evaluation of our proposed dependent variables.

The model is built up *ab initio* based on the species, reactions and processes of major importance in the context of cell susceptibility towards PTL, and parameter values such as rate constants are extracted from the literature. Thus, the model itself summarizes the current state of knowledge on the matter in a compact and descriptive manner. This type of mathematical modelling to gain insight into the underlying mechanisms of the anti-cancer effect of PTL is novel and this study is the first of its kind in the field of plasma oncology. Furthermore, we propose new ways to quantify the selective and synergistic anti-cancer effect of PTL in terms of inherent cell features, which is also an innovative approach in the ongoing research on the mode of action of PTL.

We find that the temporal maximal intracellular $H_2O_2$ concentration shows a dependency of the $H_2O_2$ membrane diffusion rate constant and the intracellular catalase concentration, such that it could possibly be used to quantify the anti-cancer effect of exogenous $H_2O_2$, but it does not account for the synergistic effect of $H_2O_2$ and $NO_2^-$ in PTL. However, by including the reactions where $CO_3^{\bullet-}$ is produced in the $CO_2$ catalyzed consumption of $ONOO^-$, and the interaction between $CO_3^{\bullet-}$ and catalase, the dependent variable $c_{1,\max}$ could possibly be able to account for the synergetic effect of $NO_2^-$ as well.

We believe that our model is an important step to unveil the underlying mechanisms of the anti-cancer effect of CAP and PTLs, but more efforts are needed in order to understand the full picture of causes and action. Here, both positive and negative results are important to share, in order to increase our collective knowledge of which clues may lead us forward in our search, and which clues we can leave behind, at least for now. Theoretical and experimental approaches to investigate possible key features of cells and their interaction with CAP and PTLs play complementary roles in our aim to push the limit of knowledge further. We hope, and believe, that our study contributes to the quest to quantify selective and synergistic effects of plasma for cancer treatment.

# Appendix A. Additional results

### A.1. Temporal maximum of the intracellular peroxynitrite concentration

Figure A1 shows the result for the dependent variable $c_{4,\max}$ for $[H_2O_2]_0^{EC} = 1\ \mu M$, with and without $NO_2^-$. The same results, but for $[H_2O_2]_0^{EC} = 1\ mM$, are shown in Figure A2.

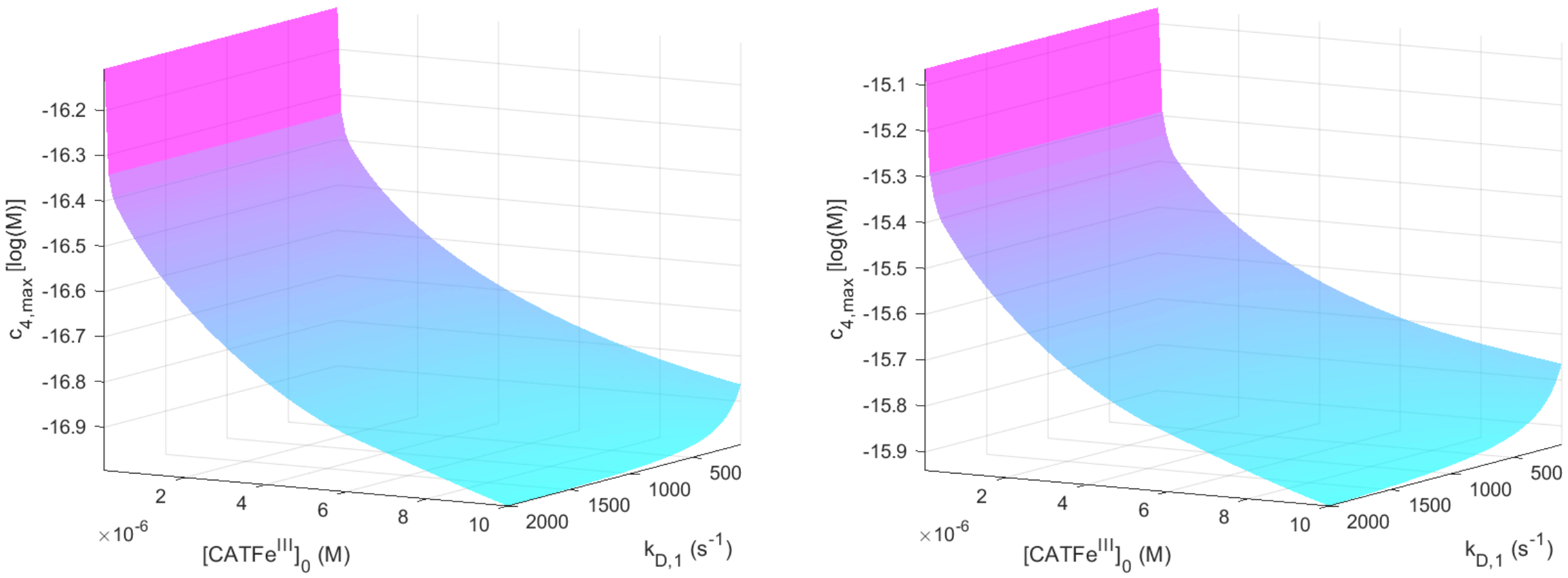

(a) (b)

**Figure A1.** The dependent variable $c_{4,\max}$ (i.e., the temporal maximum of $[ONOO^-]$ in the IC) as a function of $k_{D,1}$ and $[CATFe^{III}]_0$ when $[H_2O_2]_0^{EC} = 1\ \mu M$. $[NO_2^-]_0^{EC} = 0\ M$ (a) and $[NO_2^-]_0^{EC} = 1\ mM$ (b).

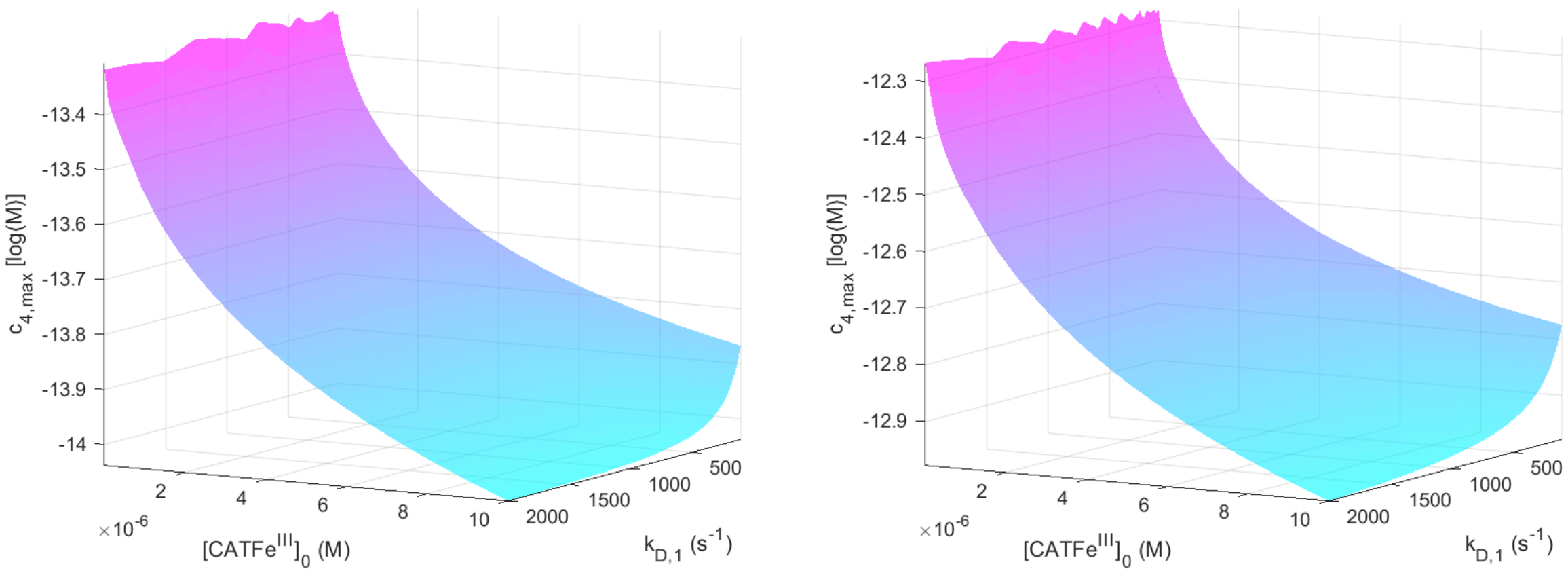

(a) (b)

**Figure A2.** The dependent variable $c_{4,\max}$ (i.e., the temporal maximum of $[ONOO^-]$ in the IC) as a function of $k_{D,1}$ and $[CATFe^{III}]_0$ when $[H_2O_2]_0^{EC} = 1\ mM$. $[NO_2^-]_0^{EC} = 0\ M$ (a) and $[NO_2^-]_0^{EC} = 1\ mM$ (b).

The dependent variable $c_{4,\max}$ could account for selectivity with respect to the different regimes of $[H_2O_2]_0^{EC}$ in the same manner as $c_{1,\max}$, see Figures A1.a and A2.a ($c_{4,\max}$) and Figures 2.a and 3.a ($c_{1,\max}$). Indeed, there is a difference in $c_{4,\max}$ of about three orders of magnitude for $[H_2O_2]_0^{EC} = 1\ mM$ (Figure A2.a) as compared with $[H_2O_2]_0^{EC} = 1\ \mu M$ (Figure A1.a).

When comparing Figures A1.a and b, we see that although the overall behavior of $c_{4,\max}$ is very similar, there is an order of magnitude difference in its value. In other words, addition of $NO_2^-$ increases the value of $c_{4,\max}$ for all $k_{D,1}$ and $[CATFe^{III}]_0$. Thus, $c_{4,\max}$ could account for the observed synergetic effect of PTL.

We see that for $[H_2O_2]_0^{EC} = 1\ \mu M$ (Figure A1.a), $c_{4,\max}$ shows a clear $[CATFe^{III}]_0$-dependence and is relatively independent of $k_{D,1}$ for low $[CATFe^{III}]_0$ However, the $k_{D,1}$-dependence gradually increases with increasing $[CATFe^{III}]_0$. In this case, the dependence is such that $c_{4,\max}$ (for a given value of $[CATFe^{III}]_0$) is inversely dependent on $k_{D,1}$. This is not consistent with the pattern we are looking for and although the formation of intracellular $ONOO^-$ could play a role in the overall cytotoxicity of PTL, we do not believe that it plays the main role. In this context, it should also be noted that even the maximal amount formed corresponds to a very low concentration.

In summary, the dependent variable $c_{4,\max}$ could account for selectivity with respect to the concentration of $H_2O_2$ as well as the synergistic effect of PTLs. Indeed, the addition of $NO_2^-$ in the extracellular compartment does increase $c_{4,\max}$ with about one order of magnitude. This could however be expected since the formation of $ONOO^-$ is directly proportional to the concentration of $NO_2^-$ and with $[NO_2^-]_0^{EC} = 10[NO_2^-]_0^{IC}$, the intracellular concentration of $NO_2^-$ will increase with about one order of magnitude compared to when no extracellular $NO_2^-$ is added. Still, $c_{4,\max}$ does not show a dependency that is consistent with a measure of the cell susceptibility towards PTL. However, as we discuss in section 4 and 7, by instead using the temporal maximum of intracellular $[CO_3^{\bullet-}]$, which is produced in the $CO_2$-catalyzed decomposition of $ONOO^-$, as a dependent variable, a feasible measure could possibly be found.

### A.2. System response time of intracellular hydrogen peroxide

Figure A3 shows the result for the dependent variable $\tau$ for $[H_2O_2]_0^{EC} = 1\ \mu M$, with and without $NO_2^-$. The same results, but for $[H_2O_2]_0^{EC} = 1\ mM$, are shown in Figure A4.

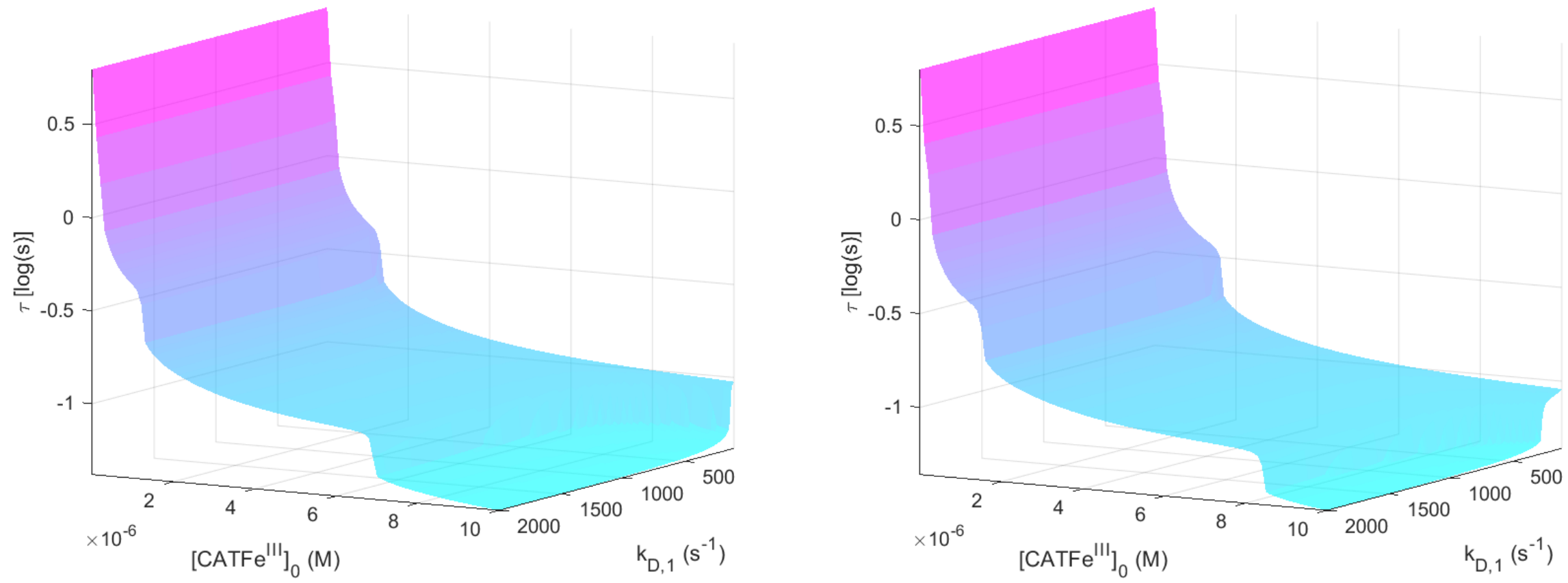

**(a)** **(b)**

**Figure A3.** The dependent variable $\tau$ (i.e., the system response time of $[H_2O_2]$ in the IC) as a function of $k_{D,1}$ and $[CATFe^{III}]_0$ when $[H_2O_2]_0^{EC} = 1\ \mu M$. $[NO_2^-]_0^{EC} = 0\ M$ (a) and $[NO_2^-]_0^{EC} = 1\ mM$ (b).

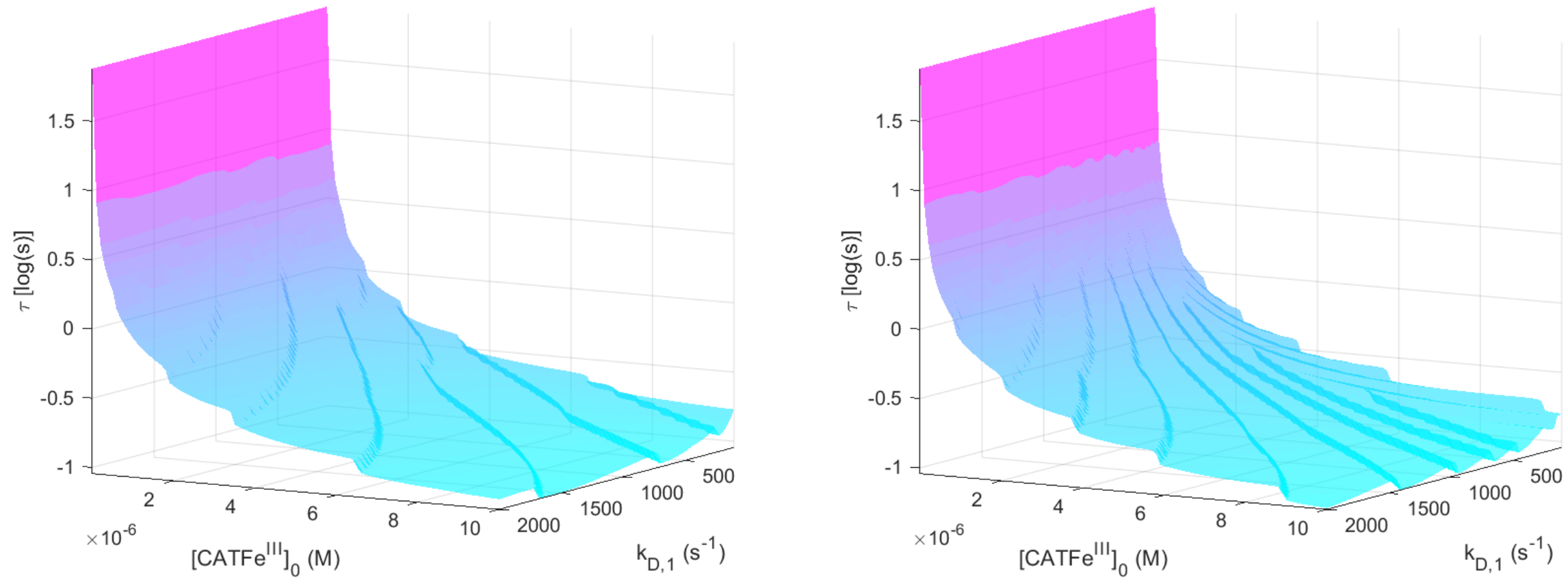

**(a)** **(b)**

**Figure A4.** The dependent variable $\tau$ (i.e., the system response time of $[H_2O_2]$ in the IC) as a function of $k_{D,1}$ and $[CATFe^{III}]_0$ when $[H_2O_2]_0^{EC} = 1\ mM$. $[NO_2^-]_0^{EC} = 0\ M$ (a) and $[NO_2^-]_0^{EC} = 1\ mM$ (b).

Here, the different concentration regimes of $[H_2O_2]_0^{EC}$ do not lead to well separated regimes of $\tau$; in Figure A3.a, we see that $-1.5 \lesssim \log(\tau) \lesssim 1$ whereas for $[H_2O_2]_0^{EC} = 1\ mM$ (Figure A4.a), $-1 \lesssim \log(\tau) \lesssim 2$. Thus, $\tau$ is not a dependent variable that clearly takes into account selectivity with respect to $[H_2O_2]_0^{EC}$. Moreover, since the system response time is decreased with increased $[H_2O_2]_0^{EC}$, cells would then be less sensitive to higher concentrations of extracellular $H_2O_2$, which is not in accordance with experimental observations.

However, there is a synergistic effect for a subspace of the total $k_{D,1}$, $[CATFe^{III}]_0$-space (see Figures A3.a and b); in the region of high $[CATFe^{III}]_0$ and for approximately the whole $k_{D,1}$-regime, the addition of $NO_2^-$ corresponds to an increased value of $\tau$.

In Figure A3.a we see that $\tau$ is more or less independent of $k_{D,1}$; the overall dominant independent variable is $[CATFe^{III}]_0$. The $[CATFe^{III}]_0$-dependence is such that the decrease of $\tau$ for increased $[CATFe^{III}]_0$ has regions with distinct drops of $\tau$ in the overall exponential decrease of $\tau$. Moreover, there is a region at high $[CATFe^{III}]_0$ and low $k_{D,1}$ where there is a slight $k_{D,1}$-dependence. This dependence is such that if a longer response time is associated with higher susceptibility, in a certain region of $k_{D,1}$, a higher value of $k_{D,1}$ has a protective effect compared to a lower value of $k_{D,1}$ (for a constant $[CATFe^{III}]_0$). This effect is increased when $NO_2^-$ is added to the system. Since this does not correspond to the current state of knowledge (see section 2.1), the system response time does not seem like a suitable dependent variable to quantify the cellular response to $H_2O_2$ and $NO_2^-$.

In summary, the system response time, $\tau$, does not seem to be a suitable measure to quantify the cell susceptibility towards PTL.

### A.3. Load of intracellular hydrogen peroxide and peroxynitrite

Figure A5 shows the result for the dependent variable $l_1$ for $[H_2O_2]_0^{EC} = 1\ \mu M$, with and without $NO_2^-$. The same results, but for $[H_2O_2]_0^{EC} = 1\ mM$, are shown in Figure A6.

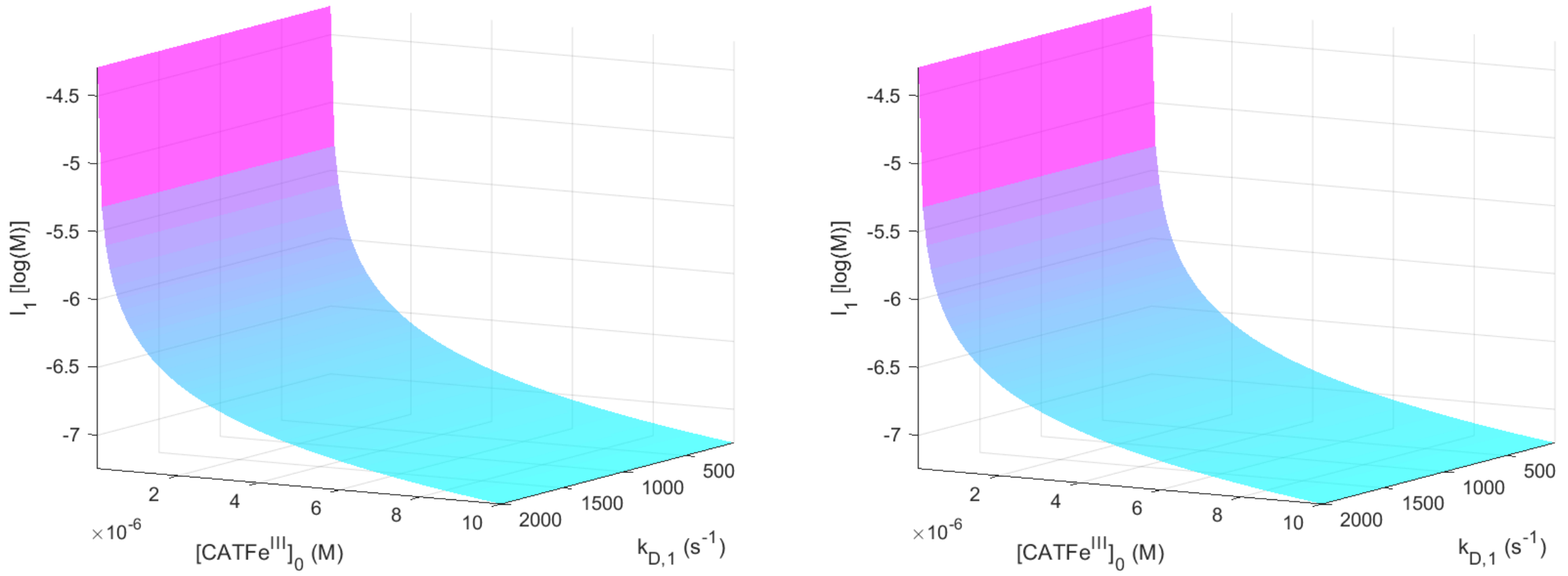

(a) (b)

**Figure A5.** The dependent variable $l_1$ (i.e., the load of $H_2O_2$ in the IC) as a function of $k_{D,1}$ and $[CATFe^{III}]_0$ when $[H_2O_2]_0^{EC} = 1\ \mu M$. $[NO_2^-]_0^{EC} = 0\ M$ (a) and $[NO_2^-]_0^{EC} = 1\ mM$ (b).

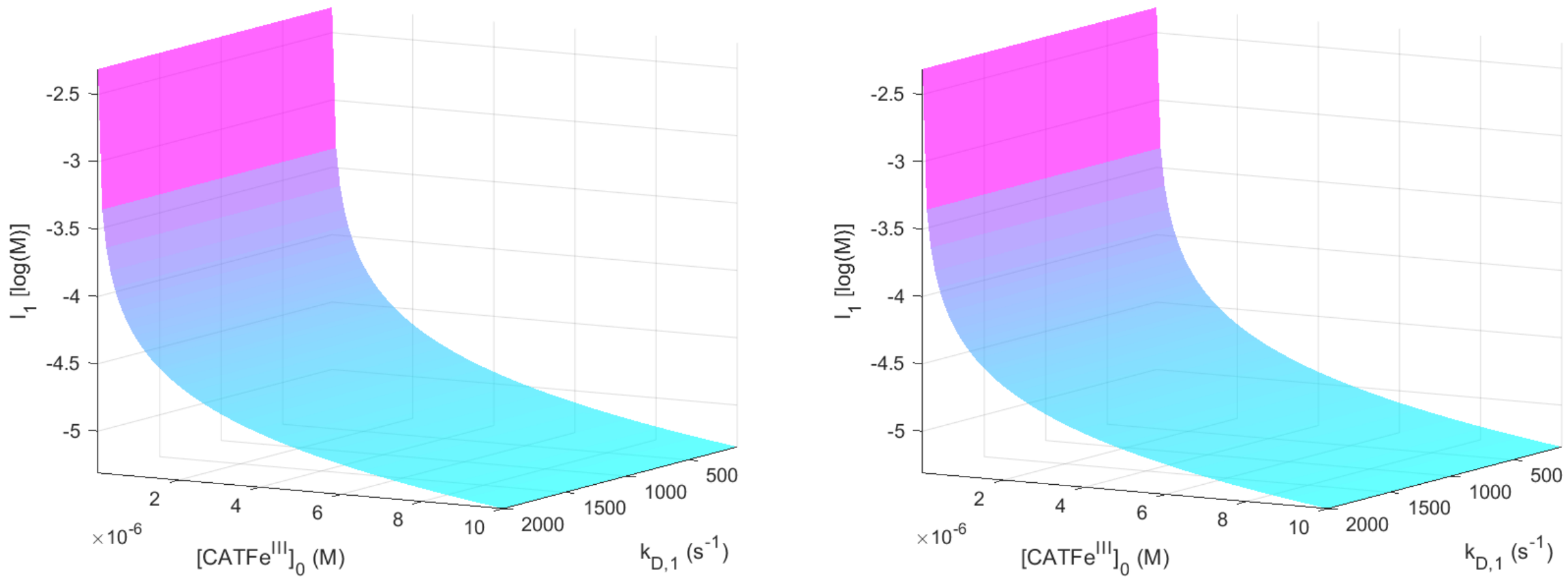

**(a)** **(b)**

**Figure A6.** The dependent variable $l_1$ (i.e., the load of $H_2O_2$ in the IC) as a function of $k_{D,1}$ and $[CATFe^{III}]_0$ when $[H_2O_2]_0^{EC} = 1\ mM$. $[NO_2^-]_0^{EC} = 0\ M$ (a) and $[NO_2^-]_0^{EC} = 1\ mM$ (b).

Likewise, Figure A7 shows the result for the dependent variable $l_{1,BS}$ for $[H_2O_2]_0^{EC} = 1\ \mu M$, with and without $NO_2^-$. The same results, but for $[H_2O_2]_0^{EC} = 1\ mM$, are shown in Figure A8.

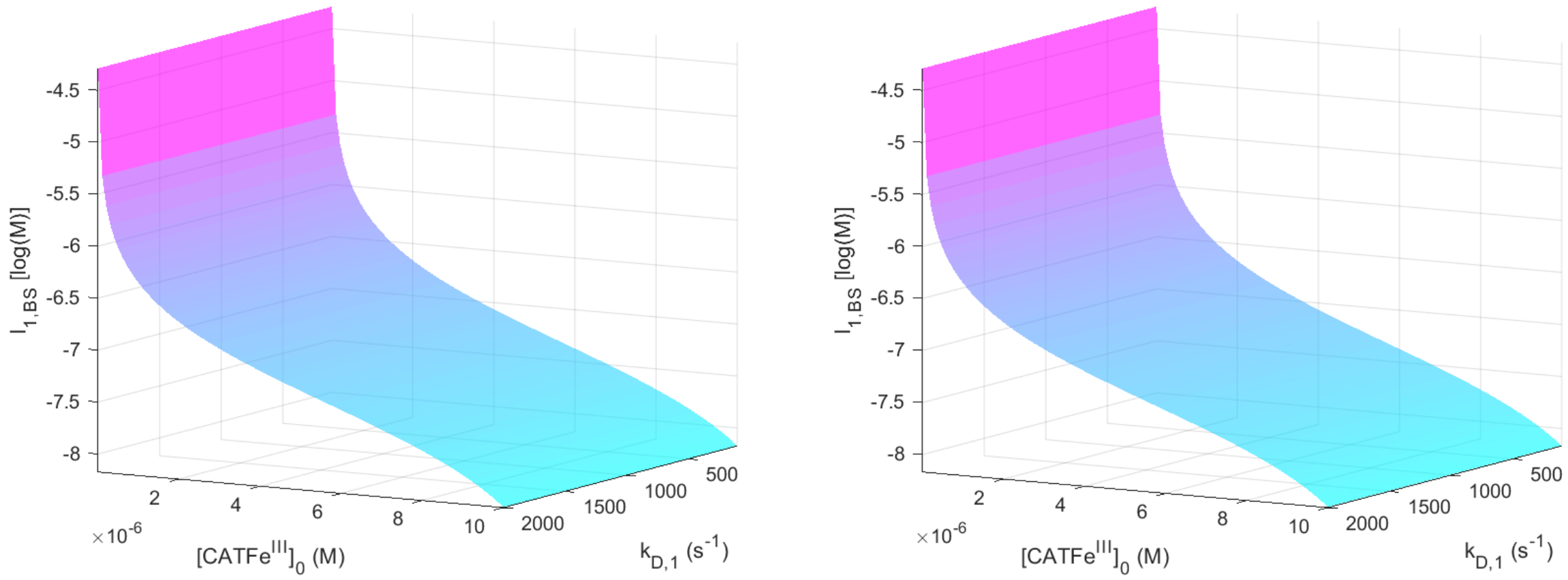

**(a)** **(b)**

**Figure A7.** The dependent variable $l_{1,BS}$ (i.e., the load over the baseline of $H_2O_2$ in the IC) as a function of $k_{D,1}$ and $[CATFe^{III}]_0$ when $[H_2O_2]_0^{EC} = 1\ \mu M$. $[NO_2^-]_0^{EC} = 0\ M$ (a) and $[NO_2^-]_0^{EC} = 1\ mM$ (b).

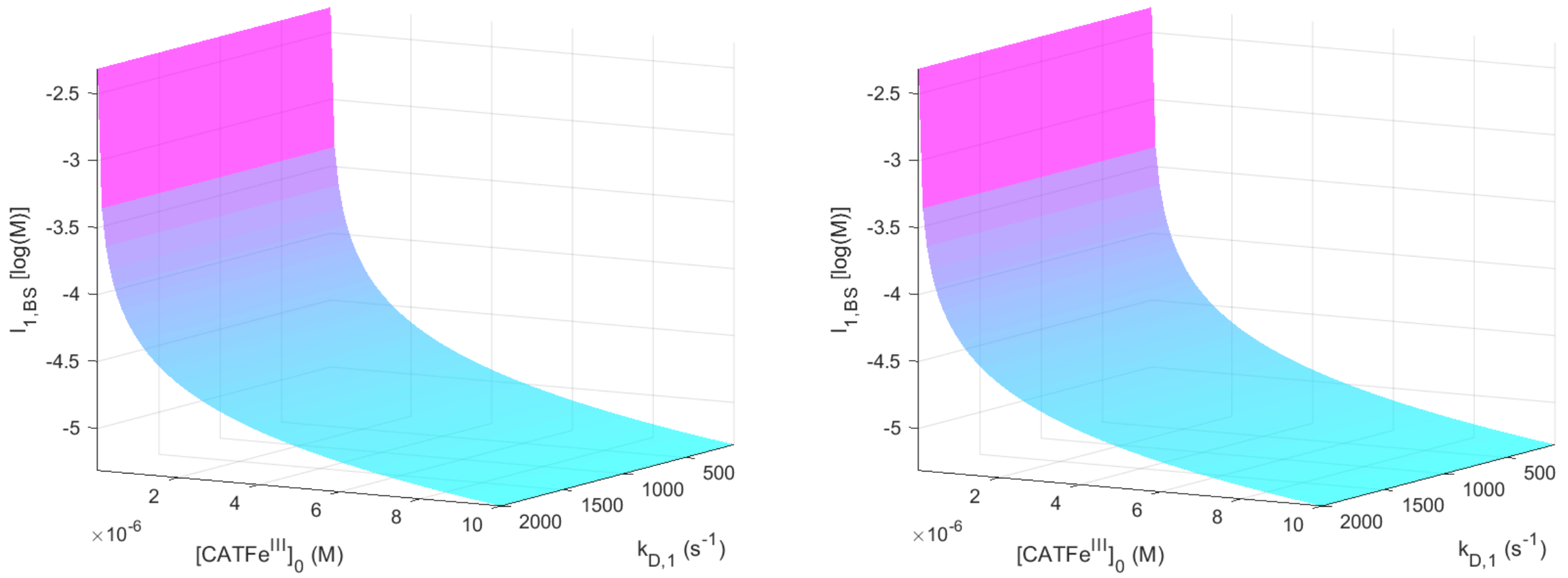

**(a)** **(b)**

**Figure A8.** The dependent variable $l_{1,BS}$ (i.e., the load over the baseline of $H_2O_2$ in the IC) as a function of $k_{D,1}$ and $[CATFe^{III}]_0$ when $[H_2O_2]_0^{EC} = 1\ mM$. $[NO_2^-]_0^{EC} = 0\ M$ (a) and $[NO_2^-]_0^{EC} = 1\ mM$ (b).

If one would exclude the systems with the lowest levels of $[CATFe^{III}]_0$, $l_1$ could account for the selectivity with respect to different $[H_2O_2]_0^{EC}$; when comparing Figures A5.a and A6.a, there is a region for which $-7.5 \lesssim \log(l_1) \lesssim -5.5$ for $[H_2O_2]_0^{EC} = 1\ \mu M$, whereas in the same region, $-5.5 \lesssim \log(l_1) \lesssim -3.5$ for $H_2O_2]_0^{EC} = 1\ mM$. The very same situation applies for $l_{1,BS}$; by excluding the lowest regime of $[CATFe^{III}]_0$, selectivity with respect to different $[H_2O_2]_0^{EC}$ could be taken into account (see Figures A7.a and A8.a).

When comparing Figure A5.a and b, there is no observable effect on $l_1$ when adding $NO_2^-$ into the system. Thus, $l_1$ does not seem to take the synergistic effect into account. The same situation applies for $l_{1,BS}$.

In Figure A5.a we see that $l_1$ is independent of $k_{D,1}$ and only dependent on $[CATFe^{III}]_0$. Thus, $l_1$ is not suitable as a measure to quantify cell susceptibility in terms of both $k_{D,1}$ and $[CATFe^{III}]_0$. From Figure A7.a we see that $l_{1,BS}$ differs a bit from $l_1$; $l_{1,BS}$ is less sensitive to an increased $[CATFe^{III}]_0$ at low $[CATFe^{III}]_0$ in particular. In addition, there is *a point of inflection* somewhere along the $[CATFe^{III}]_0$-axis, i.e., the graph is initially *concave up* and then shifts to *concave down*. This means that the rate of change of $l_{1,BS}$ with respect to $[CATFe^{III}]_0$ changes from increasing to decreasing somewhere on the $[CATFe^{III}]_0$-axis. Nevertheless, $l_{1,BS}$ is still independent of $k_{D,1}$ and thus, it does not represent a feasible measure to quantify the cell susceptibility towards PTL.

Figure A9 shows the result for the dependent variable $l_4$ for $[H_2O_2]_0^{EC} = 1\ \mu M$, with and without $NO_2^-$. The same results, but for $[H_2O_2]_0^{EC} = 1\ mM$, are shown in Figure A10.

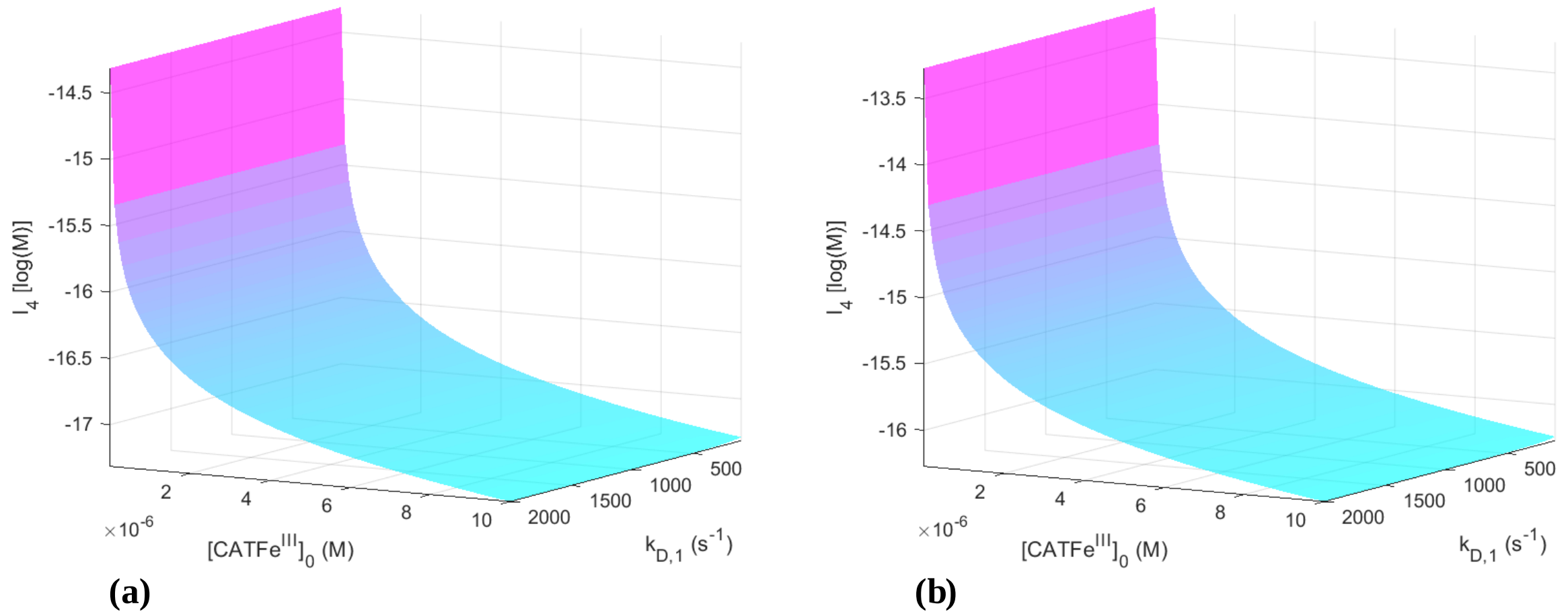

**Figure A9.** The dependent variable $l_4$ (i.e., the load of $ONOO^-$ in the IC) as a function of $k_{D,1}$ and $[CATFe^{III}]_0$ when $[H_2O_2]_0^{EC} = 1\ \mu M$. $[NO_2^-]_0^{EC} = 0\ M$ (a) and $[NO_2^-]_0^{EC} = 1\ mM$ (b).

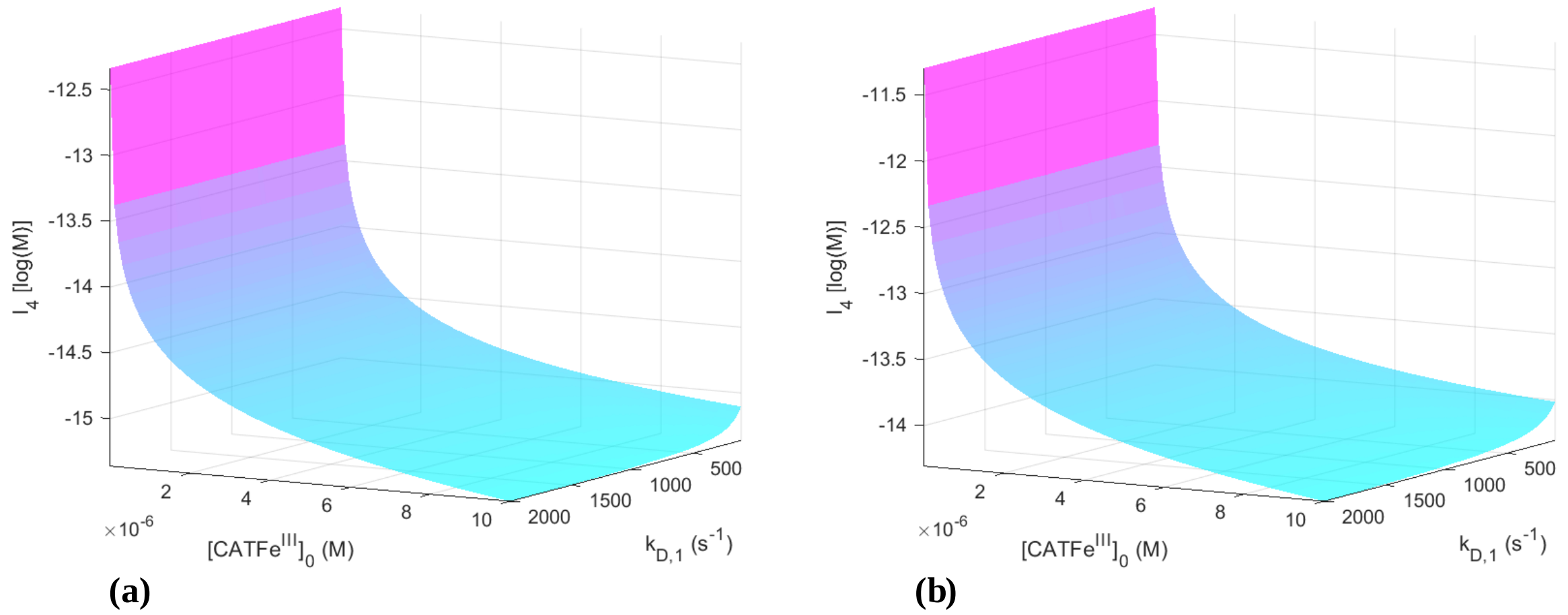

**Figure A10.** The dependent variable $l_4$ (i.e., the load of $ONOO^-$ in the IC) as a function of $k_{D,1}$ and $[CATFe^{III}]_0$ when $[H_2O_2]_0^{EC} = 1\ mM$. $[NO_2^-]_0^{EC} = 0\ M$ (a) and $[NO_2^-]_0^{EC} = 1\ mM$ (b).

The pattern for $l_4$ more or less follows the same pattern as for $l_1$; selectivity with respect to $[H_2O_2]_0^{EC}$ is only taken into account if the $[CATFe^{III}]_0$-regime is modified by removing the lowest levels of $[CATFe^{III}]_0$. In addition, $l_4$ is independent of $k_{D,1}$ and thus, it does not represent a feasible measure for quantifying cell-susceptibility. However, there exists a synergistic effect; $l_4$ is about one order of magnitude higher at every point in the $(k_{D,1},\ [CATFe^{III}]_0)$-space when $NO_2^-$ is added.

In summary, looking for a variable that depends on both $[CATFe^{III}]_0$ and $k_{D,1}$, none of the dependent variables $l_1$, $l_4$ or $l_{1,BS}$ seem to be appropriate candidates for a dependent variable able to capture and quantify the cell susceptibility towards PTL.

## A.4. Rate of extracellular hydrogen peroxide consumption

Figure A11 shows the result for the dependent variable $\bar{s}$ for $[H_2O_2]_0^{EC} = 1\ \mu M$, with and without $NO_2^-$. The same results, but for $[H_2O_2]_0^{EC} = 1\ mM$, are shown in Figure A12.

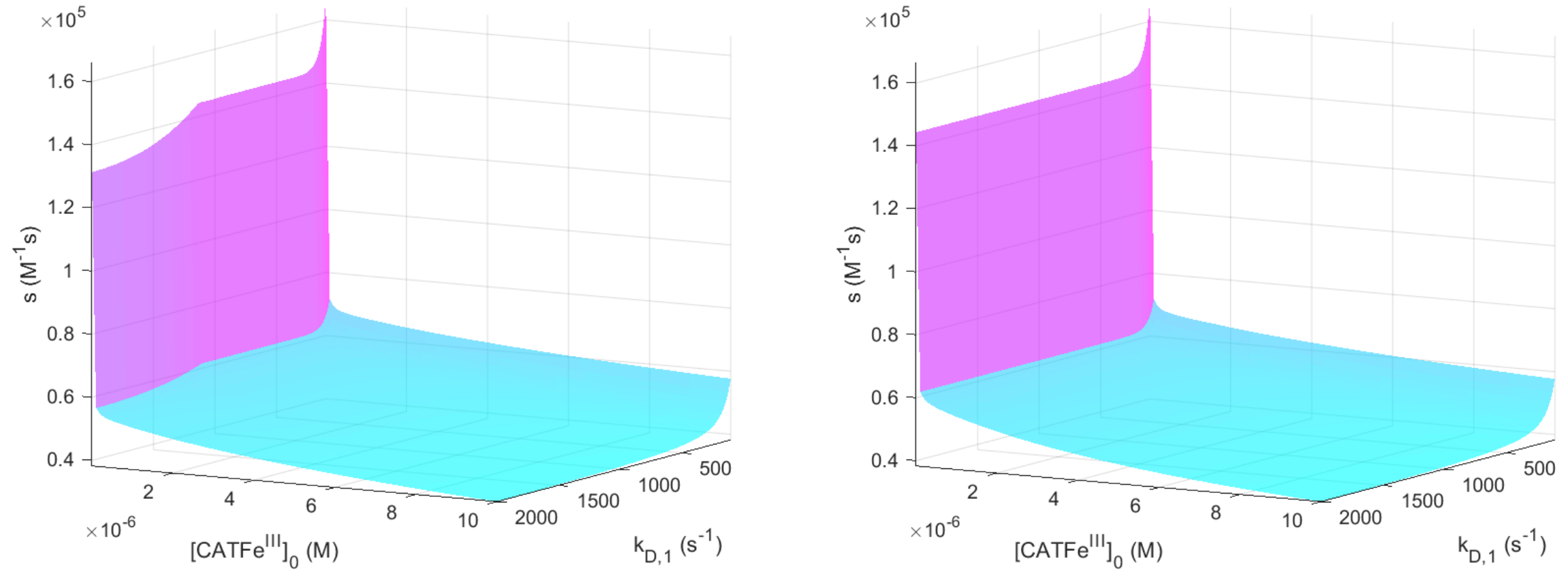

**(a)** **(b)**

**Figure A11.** The dependent variable $\bar{s}$ (i.e. the inverse of the average rate of $H_2O_2$ consumption in the EC) as a function of $k_{D,1}$ and $[CATFe^{III}]_0$ when $[H_2O_2]_0^{EC} = 1\ \mu M$. $[NO_2^-]_0^{EC} = 0\ M$ (a) and $[NO_2^-]_0^{EC} = 1\ mM$ (b).

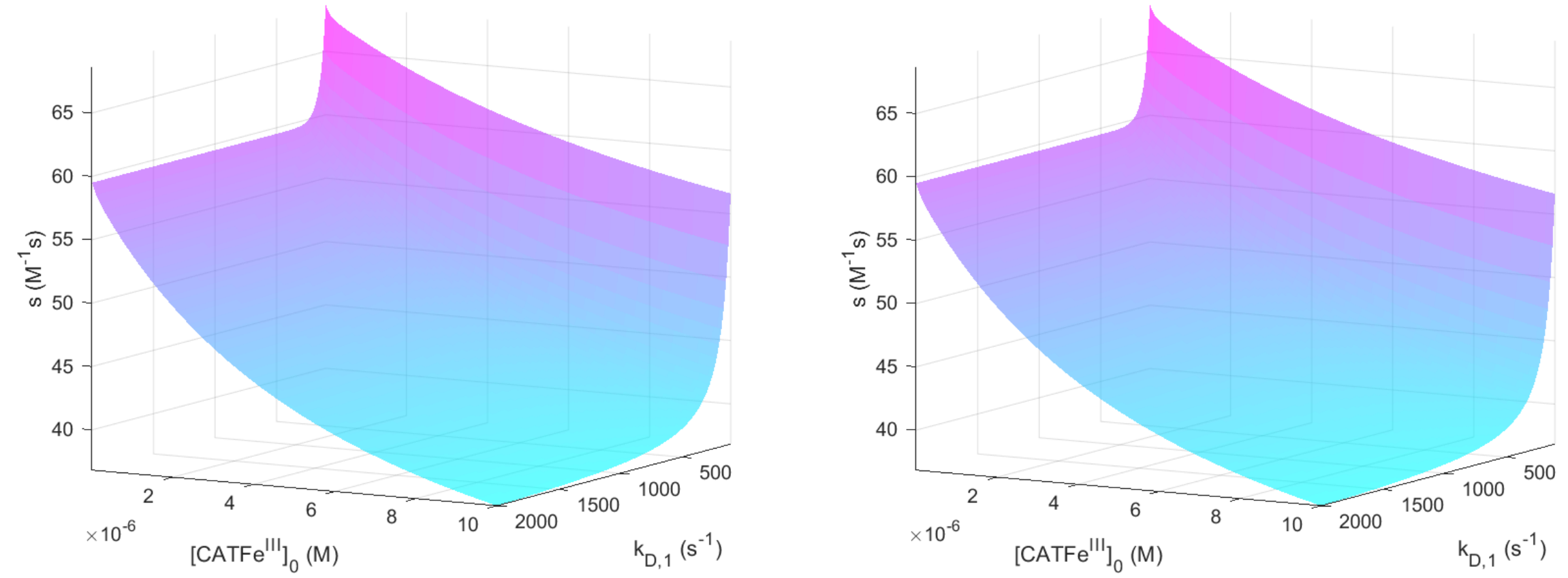

**(a)** **(b)**

**Figure A12.** The dependent variable $\bar{s}$ (i.e. the inverse of the average rate of $H_2O_2$ consumption in the EC) as a function of $k_{D,1}$ and $[CATFe^{III}]_0$ when $[H_2O_2]_0^{EC} = 1\ mM$. $[NO_2^-]_0^{EC} = 0\ M$ (a) and $[NO_2^-]_0^{EC} = 1\ mM$ (b).

Here, one has to be a bit careful to compare $\bar{s}$ for different regimes of $[H_2O_2]_0^{EC}$. Obviously, a larger value of $[H_2O_2]_0^{EC}$ will correspond to a higher membrane diffusion rate of $H_2O_2$ from the EC to the IC, which will affect the consumption rate such that it increases with increasing $H_2O_2$ membrane diffusion rate. When comparing $\bar{s}$ for $[H_2O_2]_0^{EC} = 1\ \mu M$ (Figure A11.a) and for $[H_2O_2]_0^{EC} = 1\ mM$ (Figure A12.a) we see that it is many orders of magnitude higher in the latter case, even though we know that $[H_2O_2]_0^{EC} = 1\ mM$ is associated with a higher cytotoxic effect than $[H_2O_2]_0^{EC} = 1\ \mu M$. Thus, for this choice of dependent variable, it is not meaningful to compare the results for different regimes of $[H_2O_2]_0^{EC}$; in other words, selectivity with respect to $[H_2O_2]_0^{EC}$ cannot be verified.

For the potential synergistic effect, we see in Figure A11 that even though there is a slight deviation of Figure A11.a as compared to Figure A11.b for very low values of $[CATFe^{III}]_0$, it is not enough to verify a synergistic effect when $NO_2^-$ is added to the system.

The variable $\bar{s}$ is strongly dependent on $[CATFe^{III}]_0$ in the regime of very low values of $[CATFe^{III}]_0$. Here, the dependence is such that is could capture specific cell susceptibility in terms of $[CATFe^{III}]_0$. However, for larger values of $[CATFe^{III}]_0$, $\bar{s}$ is more of less constant and moreover, the very weak $k_{D,1}$-dependence is reverse to what would be expected (see section 2.1). Thus, $\bar{s}$ does not seem to represent a feasible measure to quantify cell susceptibility in terms of $k_{D,1}$ and $[CATFe^{III}]_0$.

Finally, Figure A13 shows the result for the dependent variable $s_{\max}$ for $[H_2O_2]_0^{EC} = 1\ \mu M$, with and without $NO_2^-$. The same results, but for $[H_2O_2]_0^{EC} = 1\ mM$, are shown in Figure A14.

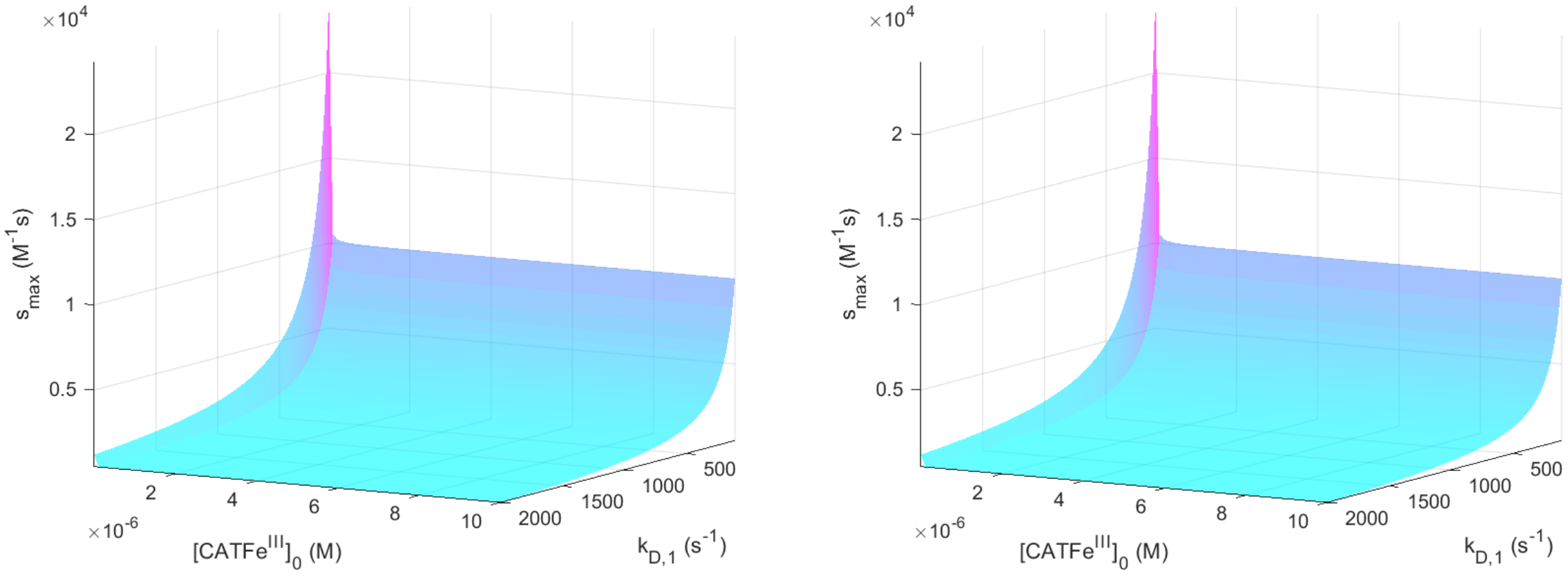

(a) (b)

**Figure A13.** The dependent variable $s_{\max}$ (i.e. the inverse of the maximal rate of $H_2O_2$ consumption in the EC) as a function of $k_{D,1}$ and $[CATFe^{III}]_0$ when $[H_2O_2]_0^{EC} = 1\ \mu M$. $[NO_2^-]_0^{EC} = 0\ M$ (a) and $[NO_2^-]_0^{EC} = 1\ mM$ (b).

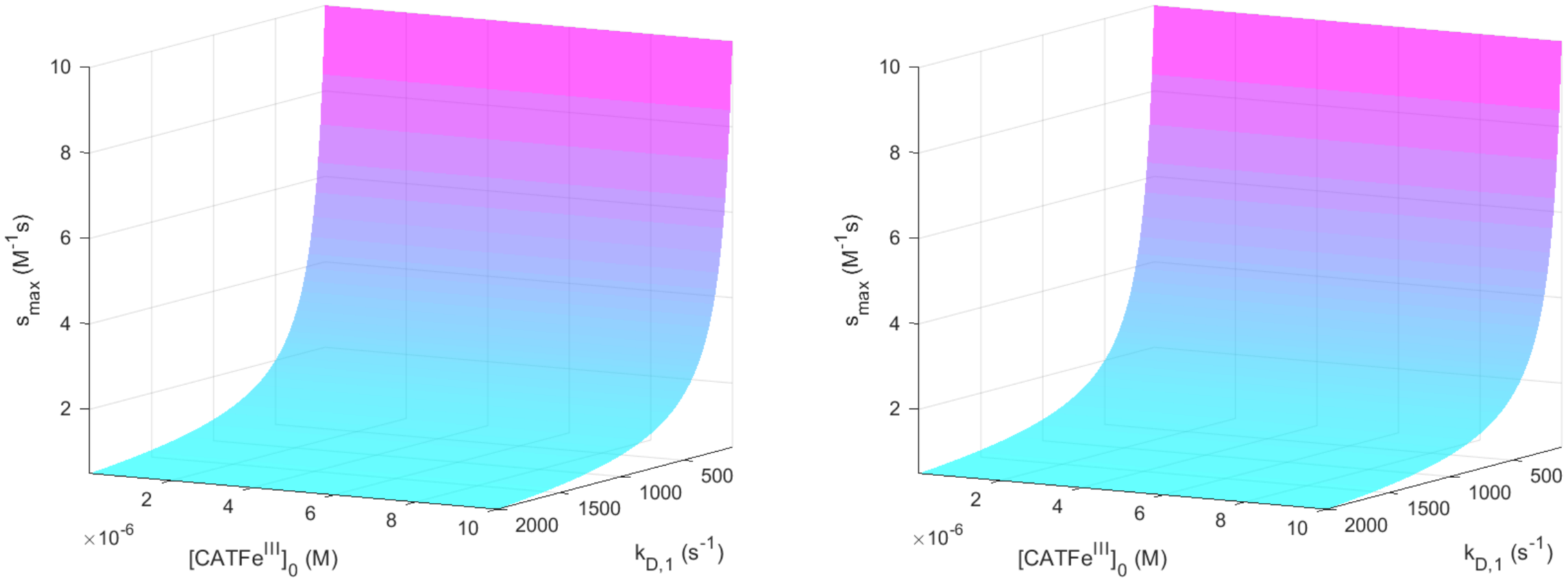

(a) (b)

**Figure A14.** The dependent variable $s_{\max}$ (i.e. the inverse of the maximal rate of $H_2O_2$ consumption in the EC) as a function of $k_{D,1}$ and $[CATFe^{III}]_0$ when $[H_2O_2]_0^{EC} = 1\ mM$. $[NO_2^-]_0^{EC} = 0\ M$ (a) and $[NO_2^-]_0^{EC} = 1\ mM$ (b).

As for $\bar{s}$, we cannot compare $s_{\max}$ for different regimes of $[H_2O_2]_0^{EC}$ (see Figures A13.a and A14.a), so again the feature of selectivity with respect to $[H_2O_2]_0^{EC}$ cannot be verified. Furthermore, there is no observable synergistic effect when $NO_2^-$ is added to the system (see Figures A13.a and b).

Opposed to $\bar{s}$, $s_{\max}$ is more or less independent of $[CATFe^{III}]_0$ and only dependent of $k_{D,1}$. However, the dependence of $k_{D,1}$ is reverse to what would be expected (see section 2.1); the lower value of $k_{D,1}$, the higher cell susceptibility towards PTL. Thus, $s_{\max}$ does not qualify as a measure to quantify the cell susceptibility in terms of $k_{D,1}$ and $[CATFe^{III}]_0$.

In summary, neither $\bar{s}$ or $s_{\max}$ seem to be appropriate candidates for a dependent variable able to capture and quantify the cell response to PTL.

## Appendix B. Derivation of the rate of diffusion equation

According to Fick's first law, the flow of a species through a membrane can be written as

$$J = \frac{1}{A}\frac{dn}{dt} = -D\frac{dc}{dx}, \tag{A1}$$

where $J$ denotes the flow in $mol\,m^{-2}s^{-1}$, $A$ is the surface area of the membrane $(m^{-2})$, $n$ is the amount of the substance $(mol)$, $D$ is the diffusion coefficient $(m^2s^{-1})$, $c$ is the concentration $(M)$. By assuming a linear concentration gradient over the cell membrane, we can write

$$\frac{dc}{dx} = \frac{c_2 - c_1}{\Delta x}, \tag{A2}$$

where $\Delta x$ is the width of the membrane, see Figure A15, Equation (A1) and (A2) yields

$$\frac{dn}{dt} = -\frac{DA}{\Delta x}(c_2 - c_1), \tag{A3}$$

$$\Leftrightarrow \ \frac{1}{V}\frac{dn}{dt} = -\frac{DA}{V\Delta x}(c_2 - c_1), \tag{A4}$$

where $V$ denotes the encapsulated volume $(m^3)$. Here, it is a custom to define the term

$$P = \frac{D}{\Delta x}, \tag{A5}$$

as the permeability of the membrane. In these terms, Equation (A4) can be written

$$\frac{dc}{dt} = \frac{-AP}{V}(c_2 - c_1). \tag{A6}$$

Thus, in our Equation (2) in the main paper,

$$k_D = \frac{-AP}{V}. \tag{A7}$$

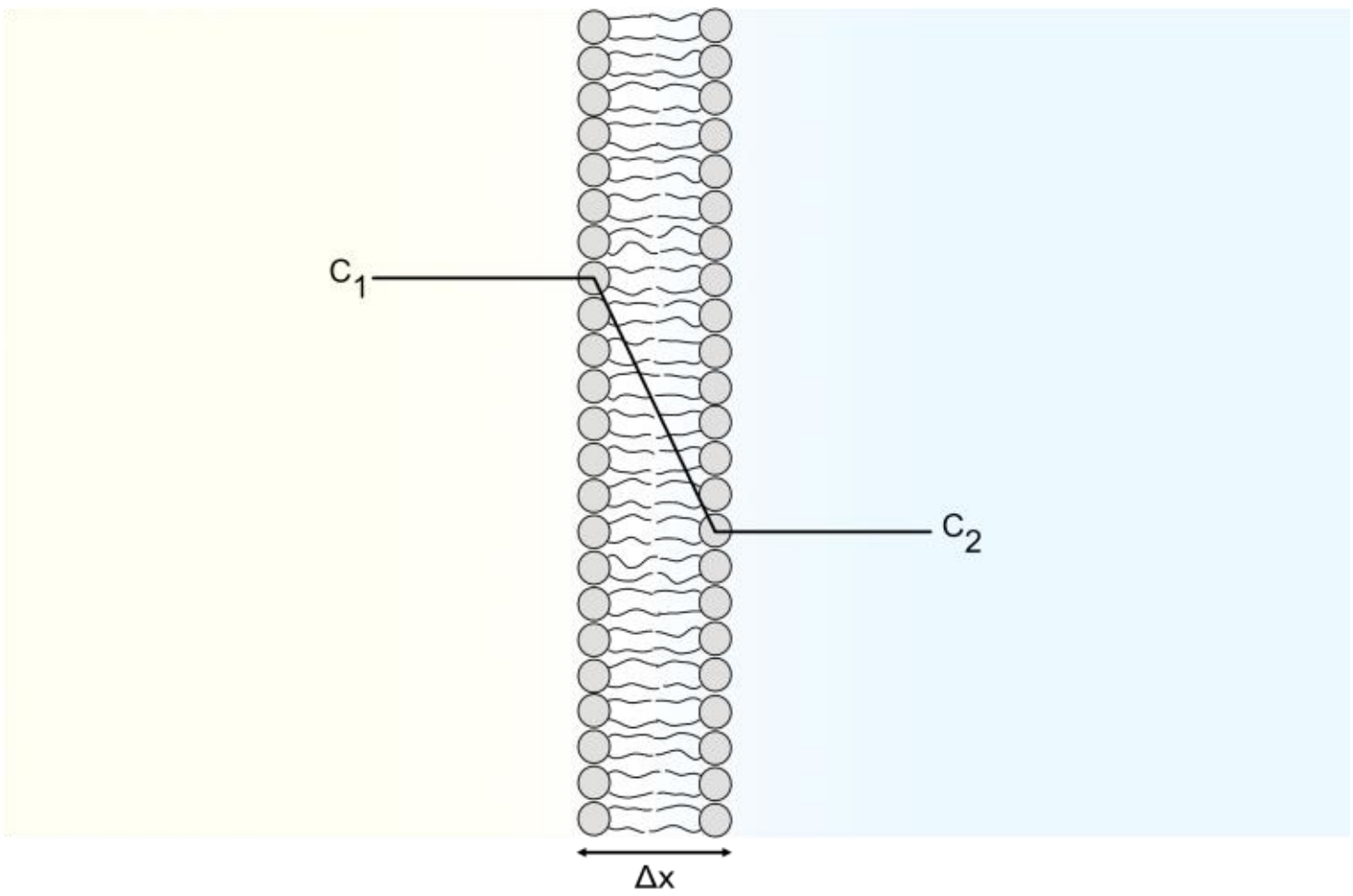

**Figure A15.** Illustration of a linear concentration gradient over a membrane.

## Appendix C. Estimates of intracellular catalase concentration

### C.1. Catalase concentration corresponding to a normal steady-state concentration of hydrogen peroxide

$H_2O_2$ is continuously produced *in vivo* [103] and remains in a quasi-steady-state. Normal intracellular steady-state concentrations of $H_2O_2$ are 10 $nM$ or less [103, 116].

Assuming that the mitochondria are the major source of intracellular $H_2O_2$, and that catalase is the only enzyme responsible for its decomposition, we estimate the average intracellular concentration of catalase. Denoting $x = [H_2O_2]$, $y = [CATFe^{III}]$, and $z = [CATFe^{IV}O^{\bullet +}]$, and including the constraint that $z = y_0 - y$, the rate equation for $x$ is

$$\frac{dx}{dt} = -x\big(k_1 y + k_2(y_0 - y)\big) + k_P. \qquad \text{(A8)}$$

Denoting the steady state concentration of $H_2O_2$ as $x_{ss}$, Equation (A8) yields

$$\begin{aligned} -x_{ss}(k_1 y + k_2(y_0 - y)) + k_P &= 0 \Rightarrow \\ y_0 &= -y\left(\frac{k_1}{k_2} - 1\right) + \frac{k_P}{k_2 x_{ss}}, \end{aligned} \qquad \text{(A9)}$$

i.e., the solutions are found on the straight line given by equation (A9). For e.g. $y = y_0$ ($x_{ss} \sim 10^{-8}$ $M$),

$$y_0 = \frac{k_P}{k_1 x_{ss}} \sim 6 \times 10^{-7}\ M.$$

### C.2. Catalase concentration from detected catalase monomers per cell

In ref. [62], the effective number of fully active catalase monomers per cell for various cancer cell lines was detected. This number varied from $101 \times 10^3$ to $538 \times 10^3$ and there was a strong correlation between the rate constant of $H_2O_2$-decomposition and the number of fully active catalase monomers per cell. Since each catalase molecule consists of four monomers, the number of catalase molecules per cell is thus roughly $(25 - 125) \times 10^3$. The conclusions in ref. [62] were that the rate constant for removal of extracellular $H_2O_2$ are on average two times higher in normal cells than in cancer cells, and that catalase activity is critical in removing this $H_2O_2$. If normal cells have a capacity that is twice are large to remove $H_2O_2$, the number of catalase molecules for the cancer cell lines normal counterparts should be $(50 - 250) \times 10^3$. If $N$ denotes the number of molecules per cell, the number of moles per cell is

$$n = \frac{N}{N_A},$$

where

$$N_A \sim 6.022 \times 10^{23}\ mol^{-1},$$

is Avogadro's number. If each cell can be assumed to be a sphere of radius $r \sim 20\ \mu M$ (as has been estimated for HeLa-cells [117]) its volume is given by

$$V = \frac{4\pi}{3} r^3.$$

Thus, the average concentration of catalase per cell is

$$[CATFe^{III}] = \frac{n}{V}.$$

For $N = 100 \times 10^3$, $[CATFe^{III}] \sim 6 \times 10^{-6}\ M$.

## Appendix D. Concentration of intracellular nitrite and carbon dioxide

The $NO_2^-$-levels measured in human physiological fluids are $0.5 - 210\ \mu M$ [111-113].

The partial pressure in human alveolar has been found to be $p_{CO_2} = (4.0 - 9.3) \times 10^3\ Pa$ [110], which according to Henry's law yields $[CO_2] = Hp$, where $H = 3.4 \times 10^{-2} M\ atm^{-1}$ is the Henry's constant for $CO_2$ in water at $T = 298.15\ K$, i.e., $[CO_2] = (1.3 - 3.1) \times 10^{-3}\ M$. Cell culture media (Eagle's) contain $2200\ mgl^{-1}$ of $NaHCO_3$, i.e. $[HCO_3^-] = 36\ mM$. At $pH = 7.0$, this corresponds to $[CO_2] = 36\ mM$.